\documentclass[aps,prb,twocolumn,superscriptaddress]{revtex4-1}
\def\be{\begin{equation}}
\def\ee{\end{equation}}
\def\bea{\begin{eqnarray}}          
\def\eea{\end{eqnarray}}
\def\bi{\begin{itemize}}
\def\ei{\end{itemize}}
\def\nb{\nonumber}
\usepackage{xcolor}
\usepackage{graphicx}
\usepackage{amsmath}
\usepackage[T1]{fontenc}

\usepackage[colorlinks=true,citecolor=blue,linkcolor=blue,urlcolor=blue]{hyperref}

\begin{document}

\title{Sonic horizons and causality in the phase transition dynamics}

\author{Debasis Sadhukhan} 
\affiliation{Jagiellonian University, Institute of Theoretical Physics, \L{}ojasiewicza 11, 30-348 Krak\'ow, Poland}  

\author{Aritra Sinha} 
\affiliation{Jagiellonian University, Institute of Theoretical Physics, \L{}ojasiewicza 11, 30-348 Krak\'ow, Poland}               

\author{Anna Francuz} 
\affiliation{Jagiellonian University, Institute of Theoretical Physics, \L{}ojasiewicza 11, 30-348 Krak\'ow, Poland}   

\author{Justyna Stefaniak} 
\affiliation{Jagiellonian University, Institute of Theoretical Physics, \L{}ojasiewicza 11, 30-348 Krak\'ow, Poland}

\author{Marek M. Rams}
\affiliation{Jagiellonian University, Institute of Theoretical Physics, \L{}ojasiewicza 11, 30-348 Krak\'ow, Poland}       

\author{Jacek Dziarmaga} 
\affiliation{Jagiellonian University, Institute of Theoretical Physics, \L{}ojasiewicza 11, 30-348 Krak\'ow, Poland}
         
\author{Wojciech H. Zurek}
\affiliation{Theory Division, Los Alamos National Laboratory, Los Alamos, New Mexico 87545, USA}

\date{ \today }


\begin{abstract}
A system gradually driven through a symmetry-breaking phase transition is subject to the Kibble-Zurek mechanism (KZM). As a consequence of the critical slowing down, its state cannot follow local equilibrium, and its evolution becomes non-adiabatic near the critical point. 
In the simplest approximation, that stage can be regarded as ``impulse'' where the state of the system remains unchanged. It leads to the correct KZM scaling laws. However, such ``freeze-out'' might suggest that the coherence length of the nascent order parameter remains unchanged as the critical region is traversed. By contrast, the original causality-based discussion emphasized the role of the {\it sonic horizon}: domains of the broken symmetry phase can expand with a velocity limited by the speed of the relevant sound. This effect was demonstrated in the quantum Ising chain where the dynamical exponent $z=1$ and quasiparticles excited by the transition have a fixed speed of sound. 
To elucidate the role of the sonic horizon, in this paper we study two systems with $z>1$ where the speed of sound is no longer fixed, and the fastest excited quasiparticles set the size of the sonic horizon. Their effective speed decays with the increasing transition time.
In the extreme case, the dynamical exponent $z$ can diverge such as in the Griffiths region of the random Ising chain where localization of excited quasiparticles freezes the growth of the correlation range when the critical region is traversed.
Of particular interest is an example with $z<1$ --- the long-range extended Ising chain, where there is no upper limit to the velocity of excited quasiparticles with small momenta. 
Initially, the power-law tail of the correlation function grows adiabatically, but in the non-adiabatic stage it lags behind the adiabatic evolution.    
\end{abstract}

\maketitle

\section{ Introduction } 

Kibble-Zurek mechanism (KZM) evolved from the scenario for defect creation in cosmological symmetry-breaking phase transitions \cite{K-a, *K-b, *K-c}. As the post-Big-Bang Universe cools, causally disconnected regions must choose broken symmetry vacuum independently. Such random choices lead to topologically nontrivial configurations that survive phase ordering as topological defects. In the cosmological setting average size of the causally connected regions (hence, the average density of defects) is set by the Hubble radius at the time of the transition. This early Universe scenario relies on the speed of light and does not apply to the laboratory phase transitions. However, it was the point of departure for the dynamical theory \cite{Z-a, *Z-b, *Z-c, Z-d} that employs critical exponents of the transition and the quench time to predict the scaling of the resulting density of defects. KZM was successfully tested using numerical simulations \cite{KZnum-a,KZnum-b,KZnum-c,KZnum-d,KZnum-e,KZnum-f,KZnum-g,KZnum-h,KZnum-i,KZnum-j,KZnum-k,KZnum-l,KZnum-m} and laboratory experiments in condensed matter systems \cite{KZexp-a,KZexp-b,KZexp-c,KZexp-d,KZexp-e,KZexp-f,KZexp-g,KZexp-gg,KZexp-h,KZexp-i,KZexp-j,KZexp-k,KZexp-l,KZexp-m,KZexp-n,KZexp-o,KZexp-p,KZexp-q,KZexp-r,KZexp-s,KZexp-t,KZexp-u,KZexp-v,KZexp-w,KZexp-x}. More recently, KZM was adapted to quantum phase transitions \cite{Polkovnikov2005,QKZ1,QKZ2,d2005,d2010-a, d2010-b}. Theoretical developments \cite{QKZteor-a,QKZteor-b,QKZteor-c,QKZteor-d,QKZteor-e,QKZteor-f,QKZteor-g,QKZteor-h,QKZteor-i,QKZteor-j,QKZteor-k,QKZteor-l,QKZteor-m,QKZteor-n,QKZteor-o,QKZteor-p,QKZteor-q,QKZteor-r,QKZteor-s,QKZteor-t} and experimental tests \cite{QKZexp-a, QKZexp-b, QKZexp-c, QKZexp-d, QKZexp-e, QKZexp-f, QKZexp-g,deMarco2,Lukin18} of quantum KZM (QKZM) followed. The recent experiment \cite{Lukin18}, where a quantum Ising chain in the transverse field is emulated using Rydberg atoms, is fully consistent with the predicted scaling \cite{QKZ2,d2005}. 

The KZM is often presented in its cartoon version where -- due to the critical slowing down / closing of the energy gap -- the dynamics of the system literally freezes-out in the neighborhood of the critical point. Today, as the experiments are able not only to count the final number of defects but can also monitor and probe the state of the system during the transition, it is timely to re-investigate the causally limited spreading of correlations during the putative ``freeze-out'' stage of the evolution. 

\begin{figure}[t]
\vspace{-0.5cm}
\includegraphics[width=0.9\columnwidth,clip=true]{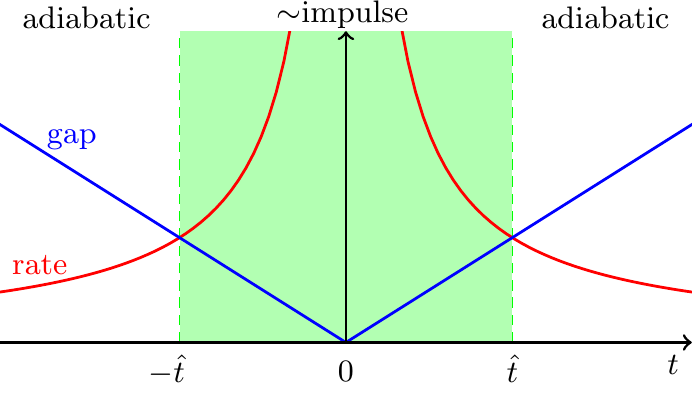}
\vspace{-0.1cm}
\caption{  
{\bf Adiabatic-impulse-adiabatic view of KZM.} 
Linear ramp crosses the critical point at time $t=0$. 
The instantaneous transition rate, $\left|\dot{\epsilon}/\epsilon\right| = 1/|t|$, diverges at the critical point and the relevant energy gap closes like $|\epsilon|^{z\nu}$. Consequently, while before $-\hat t$ the state follows the adiabatic ground state, near the critical point (between $-\hat t$ and $\hat t$) its evolution is non-adiabatic.
The freeze-out assumes that the state is ``frozen'' at $-\hat t$ 
-- size of the domains of the nascent phase does not change until $+\hat t$, where the state starts to ``catch up'' with the Hamiltonian. 
This version of KZM ignores propagation of the new phase front in the time interval $(- \hat t, +\hat t)$. It yields correct scalings, but it does not capture what happens -- for example -- in the paramagnetic-ferromagnetic quantum phase transition in the quantum Ising chain in transverse field \cite{KZscaling1,Francuzetal}. Nevertheless, it may well be relevant in phase transitions where the conserved order parameter or other causes (localization) impede propagation of phase fronts of the broken symmetry phase.
}
\label{fig:KZcartoon}
\end{figure}

In QKZM a system initially prepared in its ground state is smoothly ramped across a critical point to the other side of the quantum phase transition. A distance from the critical point, measured by a dimensionless parameter $\epsilon$ controlling a Hamiltonian, can be linearized close to the critical point as
\begin{equation}
\epsilon(t)=\frac{t}{\tau_Q}.  
\label{epsilont}
\end{equation}
Here $\tau_Q$ is a quench time. Initially, far from the critical point, the evolution is adiabatic, and the system follows its adiabatic ground state, see Fig.~\ref{fig:KZcartoon}. The adiabaticity fails at $-\hat t$ when the reaction time of the system given by the inverse of the gap becomes slower than the timescale $|\epsilon/\dot \epsilon| = |t|$ on which the transition is being imposed.
The gap closes like $\Delta\simeq|\epsilon|^{z\nu}$, where $z$ and $\nu$ are the dynamical and correlation length exponents, respectively. From the equation $|t|\simeq |t/\tau_Q|^{-z\nu}$ we obtain $\hat t\simeq \tau_Q^{z\nu/(1+z\nu)}$ and the corresponding $\hat\epsilon=\hat t/\tau_Q\simeq\tau_Q^{1/(1+z\nu)}$. In the naive ``freeze-out'' version of the impulse approximation the ground state at $-\hat\epsilon$, with a corresponding correlation length
\begin{equation}
\hat\xi \simeq \tau_Q^{\nu/(1+z\nu)},  
\label{hatxi}
\end{equation}
is expected to characterize the state of the system until $+\hat t$, when the evolution can restart. In this way, $\hat\xi$ becomes imprinted on the initial state for the final adiabatic stage of the evolution after $+\hat t$. Simplistic as it is, the adiabatic-impulse-adiabatic approximation correctly predicts the scaling of the characteristic lengthscale $\hat\xi$ and the timescale 
\be 
\hat t\simeq \hat\xi^z,
\label{hatt}
\ee 
with the critical exponents and $\tau_Q$.
They both diverge in the adiabatic limit, $\tau_Q\to\infty$, where they become the unique relevant scales in the KZ scaling ansatz~\cite{KZscaling1,KZscaling2,Francuzetal}. For instance, a two-point correlation function $C_R(t)$, between two sites separated by a distance $R$, should satisfy
\be 
\hat\xi^{\Delta} C_R(t) = F\left(t/\hat\xi^z,R/\hat\xi\right).
\label{CRscaling}
\ee
Here $\Delta$ is a scaling dimension and $F$ a non-universal scaling function. Eq.~\eqref{CRscaling} is expected to be accurate in the long-wavelength and low-frequency limit.
It is worth to observe here, that the crude adiabatic-impulse-adiabatic approximation is consistent with the scaling hypothesis \eqref{CRscaling}. However, it implies a particular  (time independent) form of the scaling function $F$.

\begin{figure}[t]
\vspace{-0.5cm}
\includegraphics[width=0.9\columnwidth,clip=true]{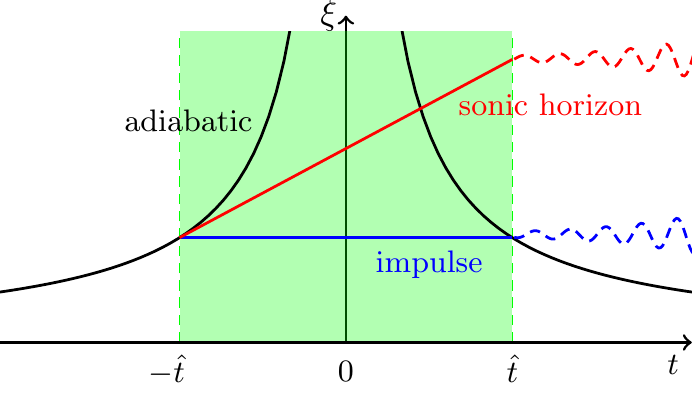}
\vspace{-0.1cm}
\caption{  
{\bf Sonic horizon view of KZM.} Initially, the correlation length $\xi$ follows adiabatically the equilibrium healing length that -- in the adiabatic ground state (black) -- diverges at the critical point. Critical slowing down means that the size of the correlation length will begin to lag behind the values dictated by the ground state of the Hamiltonian at about $- \hat t$. Pre-transition fluctuations reach size $\hat \xi$ at that instant and seed subsequent evolution of the system. The new broken symmetry phase is therefore selected by fluctuations in domains if size $\hat \xi$ at $-\hat t$. Broken symmetry spreads within the impulse time interval of $2 \hat t$ with the velocity $2\hat v$ in every direction, enlarging the resulting ``sound cone'' to roughly $ 5 \hat \xi$ by $\hat t$. In the freeze-out approximation (blue), after $-\hat t$ the correlation length freezes, and remains close to the adiabatic correlation length at $-\hat t$. Both the freeze-out and the sonic horizon views lead to the same scalings, but they result in different estimates of the pre-factors for domain sizes and defect densities.
}
\label{fig:KZreal}
\end{figure}

As emphasized already in the early papers, see Ref.~\onlinecite{Z-a, Z-b, Z-c, Z-d}, the freeze-out is not the complete story, and often not even a good approximation. A simple ``sonic horizon'' argument appealing to causality that goes beyond the impulse approximation is often more accurate. It is illustrated schematically in Fig.~\ref{fig:KZreal}. As long as the evolution is adiabatic, the rate of growth of the diverging adiabatic correlation length, $\xi\simeq|\epsilon|^{-\nu}$, is  
\be 
\frac{d\xi}{dt}=
\frac{d\epsilon}{dt}\frac{d\xi}{d\epsilon}=
\frac{1}{\tau_Q}
\frac{\nu}{|\epsilon|^{\nu+1}}. 
\ee 
This rate diverges at the critical point. Hence there must be time $-\hat t$ when it exceeds the speed limit set by twice
\be 
\hat v\simeq\frac{\hat\xi}{\hat t}\simeq\tau_Q^{-\nu(z-1)/(1+z\nu)}.
\label{hatv}
\ee 
The scaling of $-\hat t$ obtained in this way is the same as in Eq.~\eqref{hatt}. 

Causality and the KZ velocity $\hat v$ are also central for the short-cuts to adiabaticity via inhomogeneous KZM. Therein, the external driving field has a smooth position dependence, gradually taking the system across the critical point---one part after another. Velocity of the driven critical front below $\hat v$ (which in general depends on the shape of the above position dependence) is expected to pave the way to adiabatic dynamics, both for classical\cite{ inhomo_classical-a, KZnum-c, inhomo_classical-c, inhomo_classical-d, inhomo_classical-e, inhomo_classical-f} and for quantum\cite{inhomo_quantum-a, *inhomo_quantum-aa, inhomo_quantum-b, *inhomo_quantum-c, inhomo_quantum-d, inhomo_quantum-e, inhomo_quantum-f, inhomo_quantum-g} systems.

In the QKZM the speed limit is central to the causal argument. It originates from the dispersion of quasiparticles at the critical point: $\omega\simeq k^z$. Their speed for a quasimomentum $k$ is $v=d\omega/dk\simeq k^{z-1}$. Between $-\hat t$ and $\hat t$ the quench excites quasiparticles with the magnitude of $k$ up to $\hat k\simeq \hat\xi^{-1}$. The speed of quasiparticles with the largest excited $k$ is therefore $\hat v\simeq\hat k^{z-1}\simeq\hat\xi^{1-z}\simeq\hat\xi/\hat t$. When $z\geq1$, $\hat v$ is an upper bound on the velocity of quasiparticles. A quench in a translationally invariant system excites entangled pairs of quasiparticles with opposite quasimomenta: $k$ and $-k$. When moving apart, they are spreading correlations across the system. For $z\geq1$ the rate of correlation spreading is limited by twice the speed $\hat v$ of the fastest quasiparticles. 

In the crudest version, neglecting in particular dependence of $\hat v$ on the distance from the critical point, the argument implies that after $-\hat t$ the correlation range (i.e., the ``sonic cone'') continues to grow at the rate $2\hat v$ until $+\hat t$. By this time, the range increases from the initial $\hat\xi$ at $-\hat t$ to a final $\hat\xi+2\hat v \times 2\hat t\approx 5\hat\xi$. The growth is roughly five-fold. Even if the $5$ is just a very rough estimate, it shows how much the actual evolution can differ from the impulse approximation. Nevertheless, this argument confirms the role of $\hat\xi$ as the key relevant scale of length. With or without the prefactor of $5$, the final correlation length is proportional to $\hat\xi$, i.e., it scales with $\tau_Q$ in the same way as given by Eq.~\eqref{hatxi}. 

From another perspective, in the impulse approximation, the adiabatic ground state $|\psi\rangle$ corresponding to $-\hat t$ freezes out as the state of the system. After $-\hat t$ the adiabatic ground state departs from the frozen state. The frozen state becomes a superposition over adiabatic eigenbasis $|n\rangle$: $\sum_n c_n(t) |n\rangle$, where $c_n(t)=\langle n|\psi\rangle$. As a first step beyond the impulse approximation, we can include approximate dynamical phases: $\sum_n c_n(t) e^{-i\omega_n t}|n\rangle$, where $\omega_n$ is the adiabatic eigenfrequency at the critical point. In a (non-interacting) translationally invariant system, the eigenstates consist of pairs of excited quasiparticles, $|k,-k\rangle$, and the eigenfrequencies are sums of $2\omega_k$. The dynamical phase factors become scrambled -- and the phases begin to appear random -- when the largest of them, $2\omega_{\hat k}t\propto\hat k^z t$, becomes comparable to $1$ near $\hat t$. The dephasing begins when the evolution crosses over from the non-adiabatic KZ stage to the post-KZ adiabatic stage. That is when the phases definitely can no longer be ignored, but even before the cross-over the phase factors $e^{-2i\omega_k t}$ make the quasiparticle phase fronts propagate and let the quasiparticles spread the correlations across the system.

\begin{figure}[t]
\vspace{-0.1cm}
\includegraphics[width=\columnwidth]{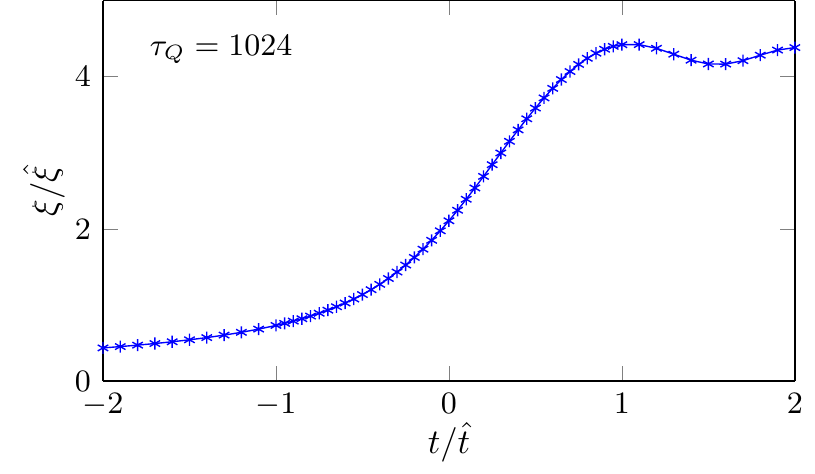}
\vspace{-0.7cm}
\caption{ {\bf Quantum Ising chain.} 
The correlation length during the quench in the 1D quantum Ising model, $H=-\sum_n(1-\epsilon) \sigma_n^z + \sigma^x_n \sigma^x_{n+1} $, where the dynamical critical exponent $z=1$ and the excited quasiparticles posses definite speed of sound. Compare with Fig.~\ref{fig:KZreal}.
Data from Ref.~\onlinecite{Francuzetal}.
}
\label{fig:Ising}
\end{figure}

For $z=1$, when the dispersion is linear in $k$ and the quasiparticles have a definite speed of sound. This effect was termed the quasiparticle event horizon \cite{EventHorizon}. In the QKZM context, it was considered in Refs. \onlinecite{KZscaling1,Francuzetal}---see Fig.~\ref{fig:Ising} for an example of the
prototypical 1D quantum Ising model.

In this paper we go beyond $z=1$ and present two examples with $z>1$: the classical Ising model with Glauber dynamics in Section \ref{sec:Onsager} and the generalized quantum XY chain in Section \ref{sec:XY}. They both exhibit an effective event horizon with a speed limit that depends on the quench time $\tau_Q$.
The generic scenario is delimited by two examples where the sonic horizon effect cease to manifest because one of its underlying assumptions is not satisfied. The first one is the random Ising model in Section \ref{sec:Random}, where localization of excited quasiparticles prevents the spreading of correlations, thus yielding in effect a ``freeze-out''. The other is the extended Ising model with long-range interactions in Section \ref{sec:LR}, where the dynamical exponent $z$ is less than $1$. The excited quasiparticles with $k\to0$ have infinite velocity, and the speed $\hat v$ at the maximal excited $\hat k$ is not an upper but a lower speed limit. Consequently, there is no sonic horizon effect, and the correlations have a long-range power-law tail that can evolve in time. After $-\hat t$ the tail begins to lag behind its adiabatic evolution. However, instead of completely freezing out, it continues to grow at a finite rate.

\begin{figure}[b]
\vspace{-0.5cm}
\includegraphics[width=\columnwidth]{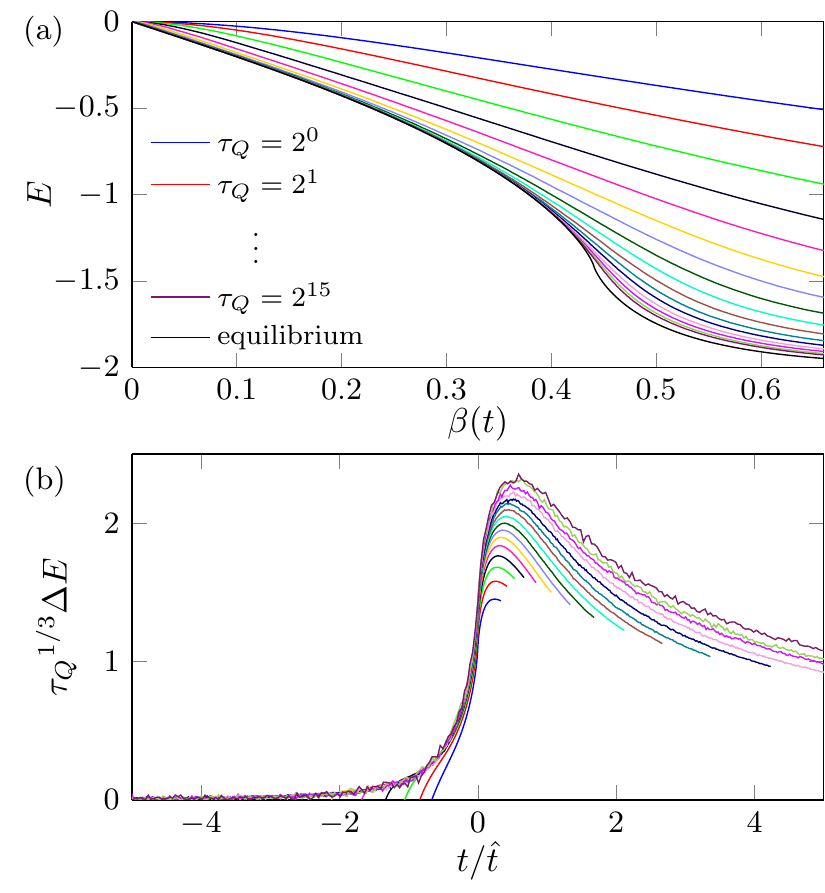}
\vspace{-0.7cm}
\caption{ {\bf Classical Ising in 2D on a square lattice.} 
In (a),
energy per site, $E$, as a function of $\beta(t)$ for different quench times $\tau_Q$.
For reference, 
the black line is the equilibrium internal energy $U(\beta)$. 
With increasing quench time, the quench curves converge to the equilibrium one.
In (b),
excitation energy per site, $\Delta E=E-U$, as a function of scaled time for different quench times $\tau_Q$.
Here, both $\Delta E$ and $t$ are rescaled according to the KZM predictions.
For large $\tau_Q\gg1$ the scaling makes them collapse in the KZ regime between 
$t\approx-\hat t$ and $t\approx\hat t$, corresponding to the rescaled times $-1$ and $+1$, respectively.
At later times the phase ordering kinetics steps in, which goes beyond the KZ physics.
}
\label{fig:onsagerenergy}
\end{figure}
\section{ Classical Ising model: ${\mathbf z=2}$ } 
\label{sec:Onsager}

We begin with the classic example of the classical Ising model on a periodic square lattice of size $L\times L$:
\be
H~=~-\sum_{\langle j,j'\rangle} \sigma^z_j \sigma^z_{j'} ~.
\label{Honsager}
\ee
On an infinite lattice, the critical inverse temperature would be $\beta_c=\ln(1+\sqrt2)/2\simeq 0.4407$.
The relevant equilibrium exponents are $\nu=1$ and $\eta=1/4$ \cite{Schultz1964}.
We model relaxation to an external heat bath with the Glauber dynamics: 
Monte Carlo update thermalizes one spin (chosen at random) at a time.
The time needed for $L^2$ such one-spin updates sets unit of time.   
For such simple relaxation, the dynamical exponent is $z=2$, belonging to the universality class of model-A dynamics~\cite{hohenberg1977theory}. 
We performed all our numerical simulations on a $4096\times 4096$ lattice. This lattice size is $100$ times longer than the longest correlation range encountered in the simulations, hence any finite size effects are eliminated with a wide safety margin. All results were averaged over $50$ repetitions of the quench, each of them starting from a different initial random spin configuration at infinite temperature.

\begin{figure*}[t]
\vspace{-0.1cm}
\includegraphics[width=\textwidth]{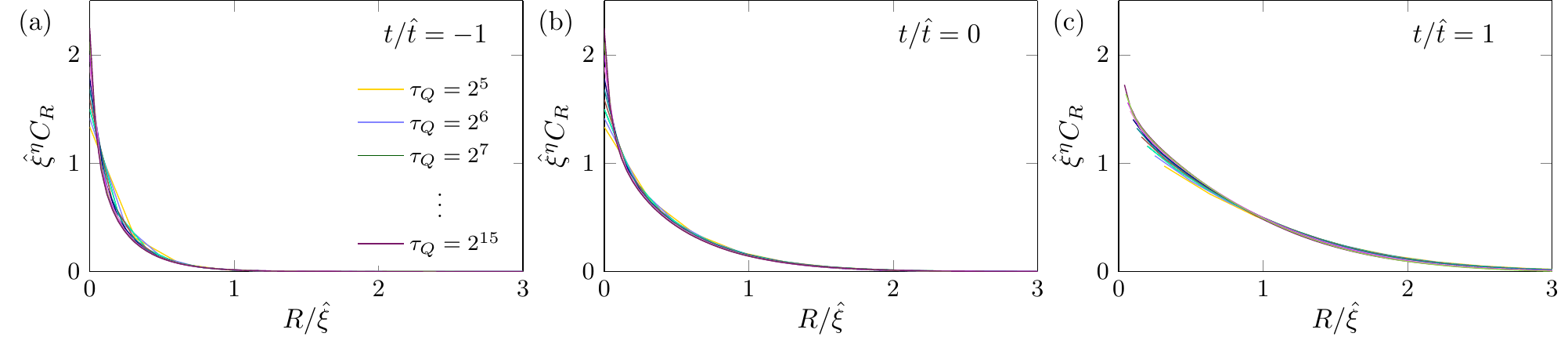}
\vspace{-0.5cm}
\caption{  
{\bf Classical Ising model in 2D.} The scaled correlator $\hat\xi^{\eta}C_R$ as a function of the scaled distance $R/\hat\xi$
for scaled times $t/\hat t=-1,0,1$ (left to right) and different quench times $\tau_Q$.
For each scaled time,
when $\tau_Q\gg1$ the plots with different $\tau_Q$ collapse to a single scaling function $F(t/\hat t,R/\hat\xi)$
demonstrating the KZ scaling (\ref{CRscaling}) hypothesis for slow enough quenches.
}
\label{fig:corrcollapse}
\end{figure*}

The inverse temperature of the heat bath is ramped linearly in time,
\be 
\beta(t)=\beta_c\left(1+\frac{t}{\tau_Q}\right),
\ee
starting with random spin configuration at $\beta(-\tau_Q)=0$. Figure~\ref{fig:onsagerenergy}(a) shows the energy $E$ during the ramp as a function of $\beta$ for different quench times $\tau_Q$. Generally, the system is more ordered for slower quenches. For slow enough quenches, the KZ picture emerges. The system evolves adiabatically until it begins to go out of equilibrium around $-\hat t$, where $\hat t\simeq\tau_Q^{2/3}$ is the KZ timescale.

In order to see how the excitation energy should depend on the quench time, let us consider the equilibrium internal energy. Near the critical point it is
\be 
U(\beta) = - \left[ 1 + A (\beta-\beta_c) \ln\left|\beta-\beta_c\right| \right]/\tanh\beta_c,
\ee
where $A\simeq 1$ is a constant. In the adiabatic-impulse approximation the state becomes effectively frozen near $-\hat t$ when $\beta_c-\hat\beta\simeq\tau_Q^{-1/3}$. At $\beta_c$ the energies of the frozen state (i.e. the instantaneous state at $-\hat t$) and the equilibrium one differ by
\bea
\Delta E(\beta_c) &=&
U(\hat\beta)-U(\beta_c) \nonumber\\
&=&
A (\beta_c-\hat\beta) \ln\left|\hat\beta-\beta_c\right|/\tanh\beta_c \nonumber\\
&\simeq &
\tau_Q^{-1/3}\ln\left(\tau_Q/\tau_0\right).
\eea 
Here $\tau_0\simeq1$ is a constant. We can see that, up to a subleading logarithmic correction, the KZ scale of energy is $\simeq\tau_Q^{-1/3}$. Accordingly, in Figure~\ref{fig:onsagerenergy}(b) we show scaled excitation energy $\tau_Q^{1/3}(E-U)$ in function of scaled time $t/\hat t$. For $\tau_Q\gg1$ the plots for different $\tau_Q$ collapse in the KZ regime: $-1<t/\hat t<1$. There is no collapse at later times when phase ordering kinetics \cite{POK} brings in new physics beyond the KZ mechanism.
Having seen how the excitations build up, we can have a closer look at the ferromagnetic correlator
\be 
C_R(t) =
\langle \sigma^z_i \sigma^z_{j} \rangle -
\langle \sigma^z_i \rangle \langle \sigma^z_{j} \rangle.
\ee

\begin{figure}[b]
\vspace{-0.1cm}
\includegraphics[width=\columnwidth]{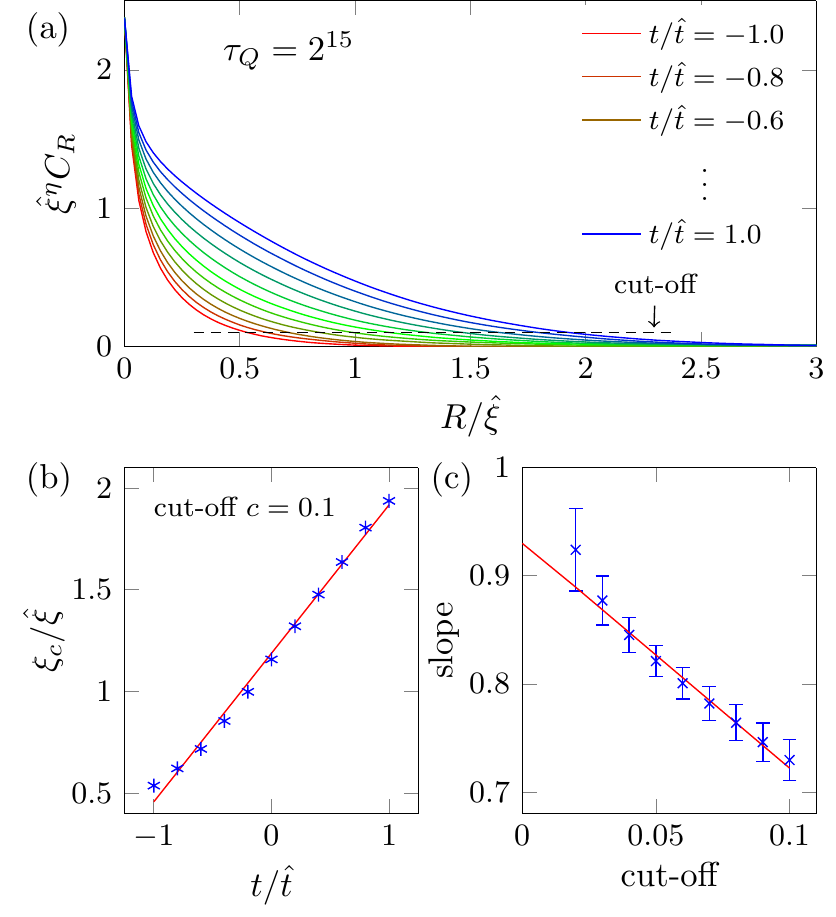}
\vspace{-0.3cm}
\caption{  {\bf Classical Ising model in 2D.}
In (a),
the scaled correlator $\hat\xi^{\eta}C_R$ as a function of the scaled distance $R/\hat\xi$ for $\tau_Q=2^{15}$ at different scaled times $t/\hat t\in[-1,1]$. The dashed line marks a cut-off $c$ (here $c=0.1$) whose crossing point is a working definition of a scaled correlation range $\xi_c/\hat\xi$.
In (b),
the scaled correlation range $\xi_c/\hat\xi$ as a function of the scaled time $t/\hat t$ for the cut-off $c=0.1$.
The best linear fit yields a slope $0.72$ as an estimate of the scaled velocity.
In (c),
the scaled velocity (slope) as a function of the cut-off. A linear extrapolation to zero cut-off yields $0.93$. This number is an estimate for (scaled) velocity at which the furthest correlation tail is spreading.
}
\label{fig:spread}
\end{figure}

\begin{figure*}[t!]
\begin{center}
\vspace{-0.1cm}
\includegraphics[width=\textwidth]{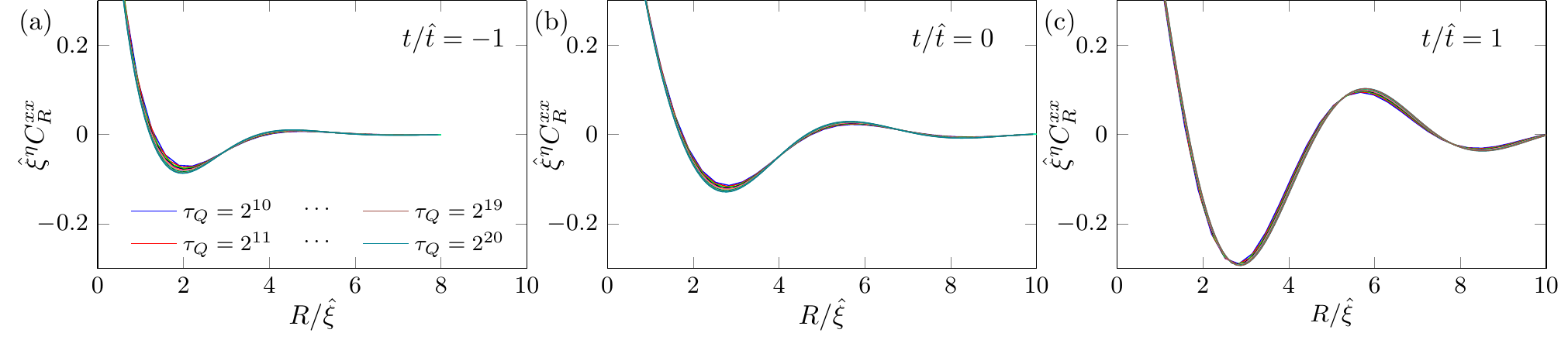}
\vspace{-0.7cm}
\end{center}
\caption{{\bf Extended quantum XY chain.} Scaled ferromagnetic correlations $C^{xx}_R$ as a function of scaled distance $R/\hat{\xi}$ for the extended XY model in (\ref{XYHsigma}) plotted for different quench times $\tau_Q$ and at different scaled times $t/\hat{t}=-1,0,1$. For large enough $\tau_Q$ plots collapse towards a universal scaling function $F(t/\hat{t},R/\hat{\xi})$ demonstrating the KZ scaling hypothesis (\ref{CRscaling}).
}
\label{fig:XYcollapse}
\end{figure*}

Here, the sites $i$ and $j$ are separated by a distance $R$ along one of the axes and the average is taken in the state at time $t$. Figure~\ref{fig:corrcollapse} shows the scaled correlator $\hat\xi^\eta C_R$ in function of the scaled distance $R/\hat\xi$ for three scaled times $t/\hat t=-1,0,1$. Here $\hat\xi=\tau_Q^{1/3}$ is the scale of length in Eq.~(\ref{hatxi}) and $\hat t=\hat\xi^z=\tau_Q^{1/3}$ is the scale of time in Eq.~(\ref{hatt}). In accordance with the scaling hypothesis (\ref{CRscaling}), for each scaled time the plots with different $\tau_Q$ collapse to a common scaling function when $\tau_Q$ is large enough. The plot for the longest $\tau_Q=2^{15}$ is practically equal to the scaling function $F(t/\hat t,R/\hat\xi)$.

Comparing the three panels in Figure~\ref{fig:corrcollapse} we can see that (at odds with the impulse approximation) the range of correlations increases several times between $-\hat t$ and $\hat t$. This observation is further corroborated by Fig.~\ref{fig:spread}(a) where we collect scaled plots for $\tau_Q=2^{15}$ at different scaled times. Using $\hat\xi$ as a natural unit of length, this figure shows how the correlation spreads in the KZ regime between $-\hat t$ and $+\hat t$. 

Figure~\ref{fig:spread}(b) shows how a scaled range of the correlator in Fig.~\ref{fig:spread}(a) depends on the scaled time $s=t/\hat t$. Here, the range is defined as the distance at which the scaled correlator falls below a threshold---set here at $0.1$. The nearly linear time dependence is fitted with a line whose slope estimates the scaled velocity of correlation spreading. We repeated the same procedure for thresholds down to $0.02$ (below $0.02$ the tail of the correlator becomes too noisy for an unambiguous estimation of the correlation range). The estimated slopes/velocities are collected in Fig.~\ref{fig:spread}(c). The velocity increases rather slowly with a decreasing threshold. In the limit of zero threshold -- probing the longest correlation tail -- it extrapolates to $0.93$. We can conclude that the speed limit in this model is $\hat v$, which allows the size of the sonic horizon to expand with
\be 
2\hat v = 
0.93~ \frac{\hat\xi}{\hat t} = 
0.93~ \tau_Q^{-1/2}.
\ee
In natural units, in accordance with KZ prediction in Eq.~\eqref{hatv}, the speed limit becomes slower for slower quenches.
\section{ Extended quantum XY chain: ${\mathbf z=3}$ } 
\label{sec:XY}

\begin{figure*}[t]
\includegraphics[width=\textwidth]{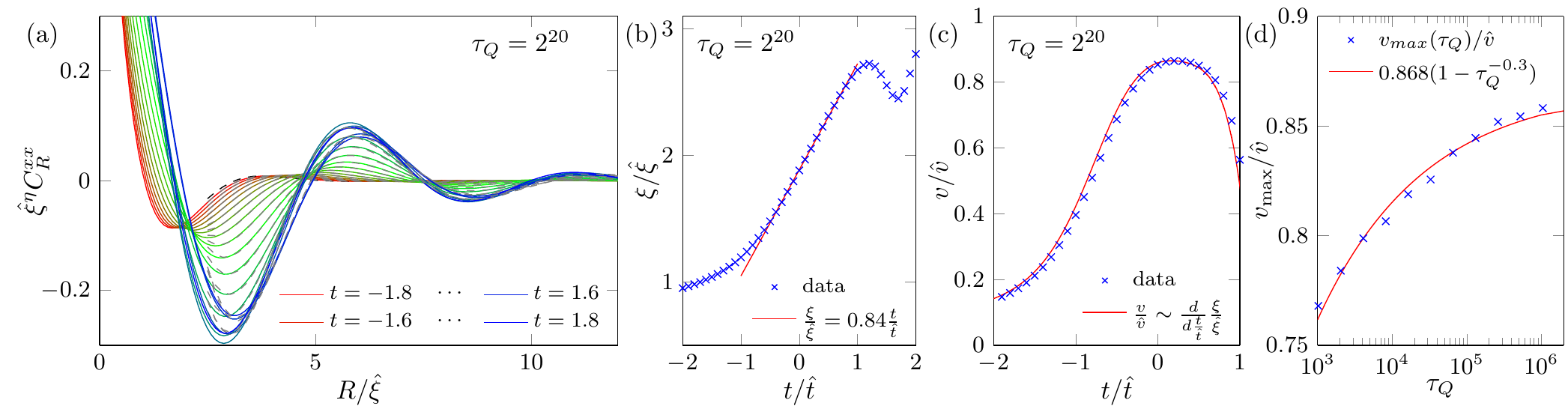}
\caption{{\bf Extended quantum XY chain.}
In (a), 
function (\ref{fitting}) is fitted (dashed lines) to the scaled ferromagnetic correlators $\hat{\xi}^{1/4}C^{xx}_R$ for different scaled times during the non-adiabatic stage of the evolution: $t/\hat{t}\in\left[-2,2\right]$.
Here $\tau_Q=2^{20}$.
In (b),
the fits provide the scaled correlation length $\xi/\hat{\xi}$ plotted against the scaled time $t/\hat{t}$. The slope of this plot is an estimate of the scaled velocity $v/\hat{v}=0.84$ at which the correlations are spreading. 
In (c),
a more accurate approximation of the velocity is obtained by taking a derivative of the plot (b) with respect to the scaled time, which yields the scaled velocity $v/\hat{v}$ as a function of the scaled time.
Maximal velocity is $v_{\rm max}/\hat v=0.84$ for $\tau_Q=2^{20}$. 
In (d),
the maximal velocity is shown as a function of $\tau_Q$.
The red line is a fit, which, for large $\tau_Q$, saturates at $v_{max}/\hat v= 0.868$. 
}
\label{fig:XYspeed}
\end{figure*}

We step into the quantum realm 
with the extended quantum XY periodic spin chain,
\bea
H&=&-\sum_{n=1}^N
\left[
(1-\epsilon) \sigma^z_n +
J^{xx}\sigma^x_n\sigma^x_{n+1}+ 
J^{yy}\sigma^y_n\sigma^y_{n+1}+
\right. \nb \\
&&
\left.
J^{xzx}\sigma^x_n \sigma^z_{n+1} \sigma^x_{n+2}+ 
J^{yzy}\sigma^y_n \sigma^z_{n+1} \sigma^y_{n+2}
\right],
\label{XYHsigma}
\eea 
a generalization of the quantum XY chain. It combines the desired $z>1$ with exact solvability.
Here we consider an anisotropic model:
\bea
&&\left(J^{xx},J^{yy},J^{xzx},J^{yzy}\right)=\nb\\
&&
\left(a\frac{1+\gamma}{2},a\frac{1-\gamma}{2},
      b\frac{1+\delta}{2},b\frac{1-\delta}{2}\right)
\eea
with the external magnetic field parametrized by $\epsilon\in[-1,1]$. The parameter is driven linearly (\ref{epsilont}) from an initial $\epsilon=-1$ in the paramagnetic phase, across the critical point at $\epsilon=0$, to a final $\epsilon=1$ in the ferromagnetic phase. After the Jordan-Wigner and Fourier transformations (\ref{JordanWigner},\ref{Fourier}), the Hamiltonian becomes:
\bea
H &=\sum_{k>0} &
A_k(\epsilon) \left(c_k^\dag c_k+c_{-k}^\dag c_{-k}\right) \nonumber \\ 
&&
+B_k \left(c_kc_{-k}+c_{-k}^\dag c_k^\dag\right),
\label{XYHk}
\eea
where,
\bea 
A_k(\epsilon) &=& 1-\epsilon-a\cos k-b\cos 2k,  \nb\\  
B_k           &=& a\gamma\sin k+b\delta\sin 2k.
\eea
It is diagonalized by eigenmodes of the stationary Bogoliubov-de Gennes equations:
\bea
\omega_k
\left(
\begin{array}{c}
U_k \\
V_k
\end{array}
\right)&=&
2\left[\sigma^z A_k(\epsilon)+\sigma^x B_k \right] 
\left(
\begin{array}{c}
U_k \\
V_k
\end{array}
\right),
\label{XYBdGk}
\eea
with eigenfrequencies
$
\omega_k=2\sqrt{A_k^2(\epsilon)+B_k^2}.
$
Here, we set $a = 4/3, b=-1/3, \gamma =1/2, \delta=1$. For the critical $\epsilon=0$, we obtain a cubic dispersion relation
\be 
\omega_k\simeq |k|^3,
\label{XYomegakz}
\ee 
expanding $\omega_k$ around $k=0$, and the dynamical exponent $z=3$. On the other hand, setting $k=0$ and expanding in small $\epsilon$, we obtain the gap opening as $\omega_0\simeq|\epsilon|^1$. Hence $z\nu=1$ and $\nu=1/3$. The dominant static ferromagnetic correlations have oscillatory tails:
\bea
C_R     &=&\langle\sigma_i^x\sigma_{i+R}^x\rangle-
           \langle\sigma_i^x\rangle\langle\sigma_{i+R}^x\rangle \nb \\
        &\sim & R^{-\eta} e^{-R/\xi}\cos(b~R/\xi+c).
\eea 
Here $\xi\sim|\epsilon|^{-\nu}$ is a static correlation length and $\eta=1/4$. Given the exponents $z$ and $\nu$, we can define the dynamical length (\ref{hatxi}) and time (\ref{hatt}) scales: $\hat\xi=\tau_Q^{1/6}$ and $\hat t=\tau_Q^{1/2}$, respectively. The fastest excited quasiparticles have velocity $\hat v\simeq\tau_Q^{-1/3}$, see Eq.~\eqref{hatv}. 

The time-dependent quench is solved in appendix \ref{XYtime} in a standard way \cite{d2005} by mapping to the Landau-Zener problem, see appendix \ref{LZ}. The KZ scaling hypothesis (\ref{CRscaling}) is demonstrated in Fig.~\ref{fig:XYcollapse} by collapse of the plots for different $\tau_Q$. A perfect collapse requires very large $\tau_Q\simeq2^{20}$, as explained by Eq.~\eqref{rescuv}. The oscillatory behaviour of $C_R$ in the original paramagnetic phase survives through the transition. In order to estimate the range of the scaled correlators, in Fig.~\ref{fig:XYcollapse} we fit their tails with oscillatory functions of the form:
\be 
\hat{\xi}^{1/4} C_R(t)=
a~(R/\hat{\xi})^{-1/4}~\mathrm{e}^{-\frac{R/\hat{\xi}}{\xi/\hat{\xi}}}\cos(b~R/\hat{\xi}+c),
\label{fitting}
\ee
where $a,b,c$ and $\xi/\hat\xi$ are the fitting parameters. We are interested how fast the scaled correlation length $\xi/\hat\xi$ grows with the scaled time $t/\hat t$. In order to reveal the universal behavior undisturbed by any short-range effects, we perform the fit in the range of scaled distances $R/\hat\xi>2.5$. Ferromagnetic correlations for different scaled times $t/\hat{t}$, together with the fits, are shown in Fig.~\ref{fig:XYspeed}(a) (for $\tau_Q=2^{20}$). The correlation length $\xi/\hat{\xi}$ as a function of scaled time is shown in Fig.~\ref{fig:XYspeed}(b). Its slope, equal to $0.838$, provides an estimate of the scaled velocity at which the correlations are spreading. For completeness, in Fig.~\ref{fig:XYspeed}(c) we plot the derivative of the plot in Fig. \ref{fig:XYspeed}(b) with respect to the scaled time. Its maximal value is shown in Fig.~\ref{fig:XYspeed}(d) as a function of $\tau_Q\in [2^{10},2^{20}]$. For large $\tau_Q$ it extrapolates to
\be 
2 \hat v = 0.868~ \frac{\hat{\xi}}{\hat{t}} = 0.868~ \tau_Q^{-1/3}.
\ee
As in the classical 2D Ising model, where also $z>1$, the correlation spreading becomes slower for slower quenches.

\section{ Random Ising model } 
\label{sec:Random}

In order to break the translational invariance and impede the propagation of excited quasiparticles that could spread correlations, next we consider the random Ising model (RIM) in one dimension defined by the Hamiltonian
\bea
H=
-\sum_{n=1}^N
\left[ (h_{n} - \epsilon)\sigma^z_n \ +
 J_{n}\sigma^x_n\sigma^x_{n+1}
\right].
\label{Hamil_RIM}
\eea 
Here both transverse fields $h_n$'s and nearest-neighbor couplings $J_n$'s are randomly selected from a uniform distribution between $0$ and $1$. The quantum critical point at $\epsilon=0$ is separating the paramagnetic phase ($\epsilon<0$) from the ferromagnetic one ($\epsilon>0$). The critical point in this model is surrounded by Griffiths region \cite{Griffiths-a, Griffiths-b, Griffiths-c}, where the presence of the so-called rare regions primarily manifests in two features: the activated dynamical scaling (the dynamical exponent $z\rightarrow\infty$) at the quantum critical point and the existence of singular regions where the linear susceptibility diverges even away from the critical point. The locally ordered rare regions act as giant spins that flip as a whole and are responsible for the exponentially slow dynamics near the critical point. These features are encapsulated in a dynamical exponent that diverges at the critical point as $z \sim|\epsilon|^{-1}$, see Ref. \onlinecite{Fisher1-a, Fisher1-b} which also shows that the correlation length exponent is $\nu = 2$. The average correlation function at criticality is a power law\cite{McCoyWu}:
\be 
C^{xx}_R=\langle \sigma_{i}^{x}\sigma_{i+R}^{x}\rangle \sim R^{-\eta},
\ee
where $\eta = \frac{3-\sqrt{5}}{2}\approx 0.38$. 

We consider ramping the parameter $\epsilon$ linearly as a function of time (\ref{epsilont}) driving the Hamiltonian from the initial paramagnetic ground state, across the critical point at $t=0$, into the ordered phase. Ref. \onlinecite{dziarmagaRIM-a, dziarmagaRIM-b, dziarmagaRIM-c} showed that the KZ correlation length $\hat \xi$ of the model varies logarithmically with the quench rate $\tau_Q$:
\bea
\centering
\hat{\xi} = \ln^2{(\tau_Q/a)},
\label{hatxiRIM}
\eea
when $\ln{(\tau_Q/a)}\gg 1$. Here $a\simeq1$ is a non-universal constant. We can see, that the dependence on $\tau_Q$ is very weak compared to any of the usual KZ power-law scalings (\ref{hatxi}). Taking into account that the correlation-length exponent near the critical point is $\nu = 2$, we get an estimate of the characteristic timescale,
\bea
\centering
\hat{t} = \frac{\tau_{Q}}{\ln{(\tau_{Q}/a)}},
\label{hattRIM}
\eea
 for $\ln{(\tau_Q/a)} \gg 1$.

\begin{figure}[t]
\includegraphics[width=\columnwidth]{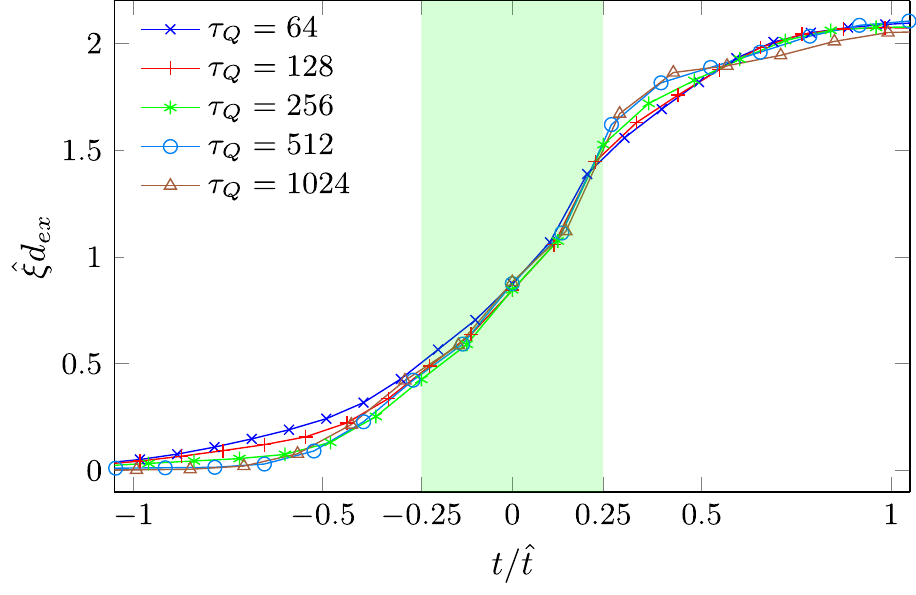}
\caption{{\bf Random Ising model.} Average scaled density of excitations $\hat\xi d_{\rm ex}$ during the quench as a function of scaled time $t/\hat t$ for different quench times $\tau_Q$. Here, we use Eqs.~(\ref{hatxiRIM},\ref{hattRIM}) with $a = 0.118$. This $a$ was tuned to obtain the best possible collapse at $t/\hat t=0$ but the plots collapse well during the whole quench. The KZ diabatic stage (shaded in green) extends roughly from $-0.25t/\hat t$ to $0.25t/\hat t$. This is where the excitation grows before it saturates in the last adiabatic stage. Averaging was done over 30 random realizations for a lattice of size $N = 128$.
}
\label{FigDoERIM}
\end{figure}
\begin{figure*}[t]
\vspace{-0.1cm}
\includegraphics[width=\textwidth,clip=true]{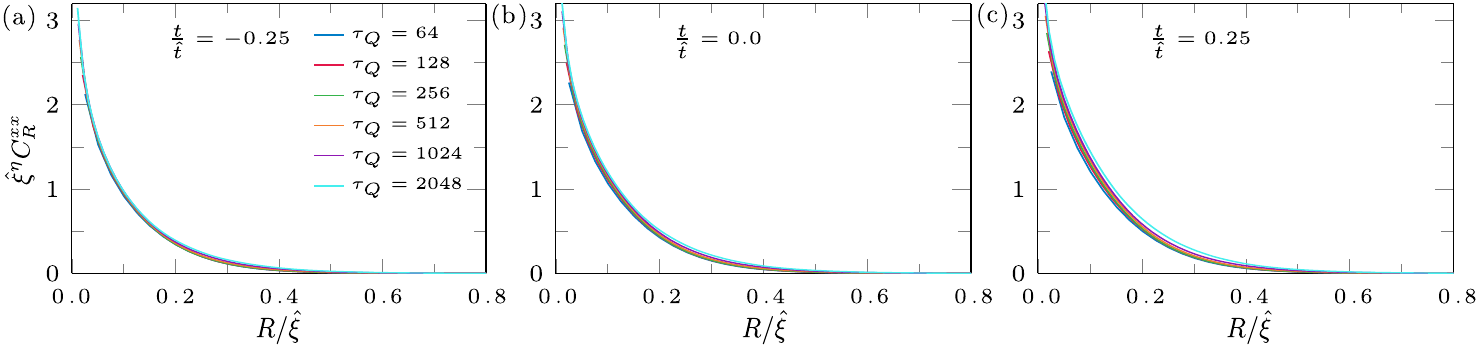}
\vspace{-0.1cm}
\caption{ {\bf Random Ising model.} Average scaled ferromagnetic correlations $\hat\xi^\eta C^{xx}_R$ as a function of scaled distance $R/\hat{\xi}$ for the random Ising model (\ref{Hamil_RIM}) plotted for different quench times $\tau_Q$ and at different scaled times $t/\hat{t}=-0.25,0,0.25$. For large enough $\tau_Q$ plots collapse towards a universal scaling function $F(t/\hat{t},R/\hat{\xi})$ demonstrating the KZ scaling hypothesis
(\ref{CRscaling}).
}
\label{fig:RIMcollapse}
\end{figure*}

In order to solve the dynamics, we map the Hamiltonian by the Jordan-Wigner transformation (\ref{JordanWigner}) to a quadratic spinless free-fermionic model
\bea
H = \sum_{n}(h_{n}-\epsilon)c_{n}^\dag c_{n} -J_{n}c_{n}^\dag c_{n+1} - J_{n}c_{n+1} c_{n} + {\rm h.c}.
\label{JW_RIM}
\eea
Following the convention of this article, we confine ourselves to the subspace of even parity of $c$-quasiparticles (anti-periodic boundary conditions, i.e., $c_{N+1} = -c_1$) and in this subspace, we diagonalize the Hamiltonian by a Bogoliubov transformation:
\bea
\centering
c_{n} = \sum\limits_{m=1}^{N}(U_{nm}\gamma_{m} + V_{nm}^* \gamma_{m}^\dag)
\label{bogoliubov_RIM}
\eea
The index $m$ labels the (Bogoliubov) eigenmodes of the stationary Bogoliubov-de Gennes equation
\bea
\centering
\omega_{m}U_{n,m}^{\pm} = 2h_{n}U_{n,m}^{\mp} - 2j_{n-1}U_{n-1,m}^{\mp},
\label{stationary_B-DeG}
\eea
where $\omega_{m}>0$, $U_{nm}^{\pm} = U_{nm}\pm V_{nm}$ and anti-periodic boundary conditions $(U_{N+1,m}^{\pm} = -U_{1,m}^{\pm},U_{0,m}^{\pm}=-U_{N,m}^{\pm})$ is implemented. The eigenstates ($U_{nm},V_{nm}$), with positive energy, $\omega_{m}>0$ normalized so that $\sum_{n}(|U_{nm}|^2 + |V_{nm}|^2) = 1$, define quasiparticle operators $\gamma_{m}=U_{nm}^{*}c_{n}+V_{nm}c_{n}^{\dag}$. We have corresponding negative energy components of the eigenstates labelled ($U_{nm}^{neg},V_{nm}^{neg}$) with energy $-\omega_{m}$, which defines a quasiparticle operator $\gamma_{m}^{neg}=(U_{nm}^{neg})^{*}c_{n}+V_{nm}^{neg}c_{n}^{\dag}$. The Bogoliubov transformation renders the Hamiltonian to be $H = \sum_{m = 1}^{N}\omega_{m}(\gamma_{m}^{\dag}\gamma_{m} - \frac{1}{2})$. In the even parity subspace only states with even number of quasi-particles belong to the spectrum of $H$.

In order to find whether the RIM fits into our narrative of the uniform scaled velocity of the correlation spreading by quasiparticles,
we proceed with the numerical simulation of the quench. 
We prepare the initial state of the system deep in the paramagnetic phase in the ground state, i.e., in the Bogoliubov vacuum state for quasiparticles at an initial time $t_{0}$ where $\epsilon(t_{0}) = 5$. As we tune the parameter $\epsilon(t)$ towards $0$, see Eq.~\eqref{epsilont}, the state of the systems departs from its adiabatic ground state and gets excited due to closing of the energy gap near the critical point. We work in the Heisenberg picture where we assume that the excited state is a Bogoliubov vacuum while the time-dependence is ascribed to a set of time-dependent quasi-particle operators
\bea
\centering
\gamma_{m}(t)=u_{nm}^{*}(t)c_{n}+v_{nm}^{*}(t)c_{n}^{\dag}
\eea
The Bogoliubov modes $u_{nm}$ and $v_{nm}$ solve the time-dependent Bogoliubov-de Gennes  equations:
\bea
\centering
i\frac{du_{nm}^{\pm}}{dt} = 2(h_{n}-\epsilon(t))u_{nm}^{\mp}-2J_{n}u_{n\mp1,m}^{\mp}
\label{t_B-deG}
\eea
We integrate equation (\ref{t_B-deG}) numerically using the 2nd order Suzuki-Trotter method, see Appendix \ref{tB-deG_RIM}. The density of excited Bogoliubov quasiparticles $d_{ex}(t)$ can be calculated at each time step by projecting the time-dependent Bogoliubov modes ($u_{nm}(t),v_{nm}(t)$) onto the corresponding instantaneous static negative Bogoliubov modes ($U_{nm}^{\text{neg}},V_{nm}^{\text{neg}}$):
\bea
\centering
d_{ex}(t) = \frac{1}{N} \sum_{s}^N\sum_{p}^N \left|\langle U_{p}^{\text{neg}},V_{p}^{\text{neg}}|u_{s}(t),v_{s}(t)\rangle\right|^2.
\label{doe_RIM}
\eea

While in the initial adiabatic stage, the system remains in the instantaneous ground state and hence the density of quasiparticle excitations is $0$. On tuning the field $\epsilon(t)$ towards its critical value at $t/\hat{t} = 0$, the system gets excited and consequently $d_{ex}$ starts to grow. As we keep increasing the field $\epsilon(t)$, $d_{ex}$ saturates as the system ends its non-adiabatic journey across the critical point. In Fig.~\ref{FigDoERIM} we see that the scaled plots (averaged over disorder) for different $\tau_Q$ collapse in accordance with the dynamical scaling hypothesis (\ref{CRscaling}). 

\begin{figure}[b]
\includegraphics[width=\columnwidth]{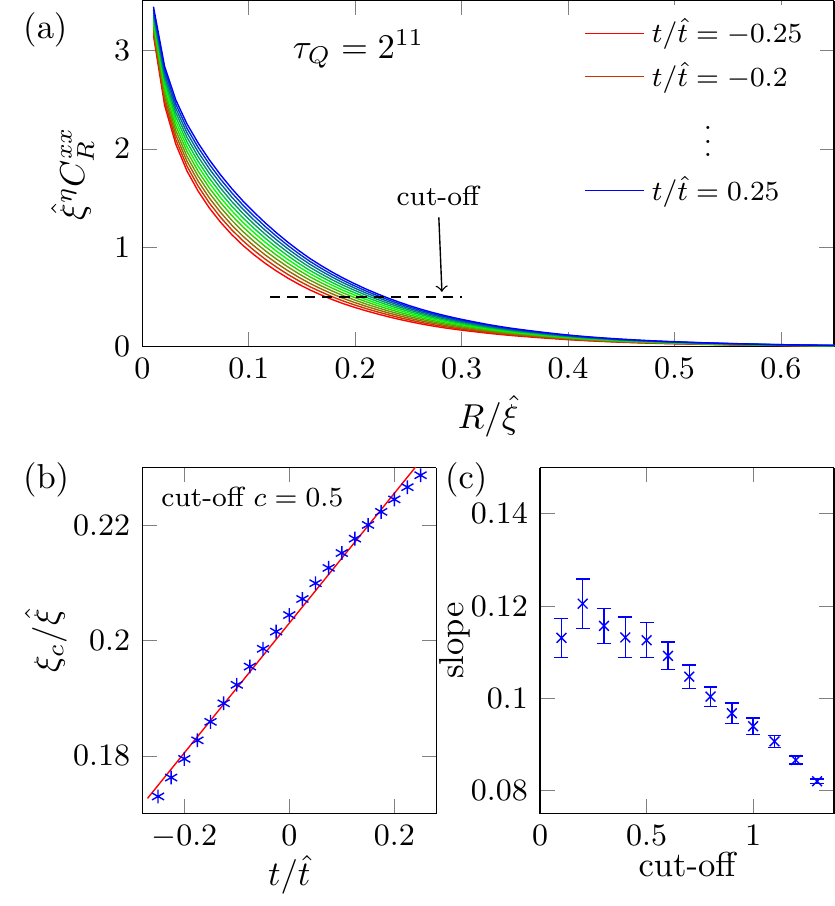}
\caption{   {\bf Random Ising model.}
In (a),
the scaled correlator $\hat\xi^{\eta}C_R$ as a function of the scaled distance $R/\hat\xi$ for $\tau_Q=2^{11}$
at different scaled times $t/\hat t$ in $[-0.25,0.25]$.
The dashed line marks a cut-off (here $1$) whose crossing point 
is a working definition of a scaled correlation range $\xi_c/\hat\xi$.
In (b),
the scaled correlation range $\xi_c/\hat\xi$ as a function of the scaled time $t/\hat{t}$ for the cut-off $c=0.5$. The best linear fit yields a slope of $0.113$ as an estimate of the scaled velocity.
In (c),
the scaled velocity (slope) as a function of the cut-off, with the error bars indicating fitting errors.
}
\label{fig:RIMspread}
\end{figure}

The hypothesis for correlations in Eq.~\eqref{CRscaling} is verified by a similar collapse of the plots in Fig.~\ref{fig:RIMcollapse}. We consider large $\tau_Q$ as the logarithmic KZ scaling laws are only valid in that regime. The correlation function is averaged over $100$ instances of disorder and lattice translations on a lattice of $N=256$. This lattice is a few times longer than the longest correlation range. The details of calculating the correlation functions are given in Appendix \ref{appcorrRIM}.

A close look at Fig.~\ref{fig:RIMcollapse} reveals that the scaled dynamical correlation function grows with scaled time. In order to appreciate the growth, we collect the scaled correlators for different scaled times in Fig.~\ref{fig:RIMspread}(a). In order to estimate the scaled correlation range $\xi_c/\hat{\xi}$ as a function of scaled time, we choose a cut-off $c$ for the scaled correlation and calculate scaled distances corresponding to the cut-off by spline interpolation. Fig.~\ref{fig:RIMspread}(b) shows the scaled range in function of the scaled time for the cut-off (selected equal to $0.5$) in Fig.~\ref{fig:RIMspread}(a).
Near the critical point, we approximate it with linear dependence where the slope is $0.113\hat\xi/\hat t$. We repeat the same procedure for other cut-offs. We present the enumerated slopes for various cut-offs in Fig.~\ref{fig:RIMspread}(c), with the error bars indicating the fitting error. Finally, we estimate the sonic horizon expansion speed in this model with the maximal observed value
\bea
\centering
2\hat{v} = 0.12 \frac{\hat{\xi}}{\hat{t}} = 0.12 \frac{\ln^3(\tau_Q/a)}{\tau_Q}
\eea
Compared to the previous models, with the prefactor $0.12$, the speed limit is significantly below expectations in this case. The entangled Bogoliubov quasiparticles get excited but, due to their localization by disorder, they do not propagate to spread correlations appreciably. In the spin language, the locally ordered regions are excited, but they essentially stay where they are. One may even argue that in this case, the adiabatic-impulse-adiabatic freeze-out approximation captures the essential physics.

\section{Long-range extended Ising model: $0<z<1$ } 
\label{sec:LR}

The long-range extended Ising model is a further generalization of the extended XY model that we have discussed in Sec. \ref{sec:XY}. Now the Hamiltonian of the system includes not only the nearest-neighbor and next-nearest-neighbor interaction, but all possible long-range cluster terms: 
\bea
H=
-\sum_{n=1}^N
\left[
 (1-\epsilon) \sigma^z_n +
\sum_{r=1}^{N-1} 
J_r\sigma^x_n \sigma^x_{n+r} \prod_{i=n+1}^{n+r-1} \sigma^z_i
\right].
\label{Hamil_LR}
\eea 
Here 
\be
J_r ~=~ \frac{1}{\zeta(\alpha)} ~ \frac{1}{r^\alpha},
\ee
and the normalization by the Riemann zeta function, possible for $\alpha>1$, is such that $\sum_r J_r=1$. The case of $\alpha>2$ is not the most interesting here because the model behaves effectively like the short-range one \cite{LRKitaev1,LRKitaev2,LR1,LR2,LRBosonic}. Similarly, when $0\le\alpha\le1$ we would need to restrict ourselves to a finite system because the thermodynamic limit does not exist in this case, and the model behaves effectively like the Lipkin-Meshkov-Glick model \cite{LMG} with infinite-range interactions. Therefore, we focus here on the intermediate $1<\alpha<2$, where a cross-over between the short and infinite-ranges happens. In this regime, we consider the linear quench (\ref{epsilont}) driving the system from the initial paramagnetic phase at $t=-\infty$, across the critical point at $t=0$, to the final ferromagnetic phase. 

\begin{figure}[b]
\includegraphics[width=\columnwidth]{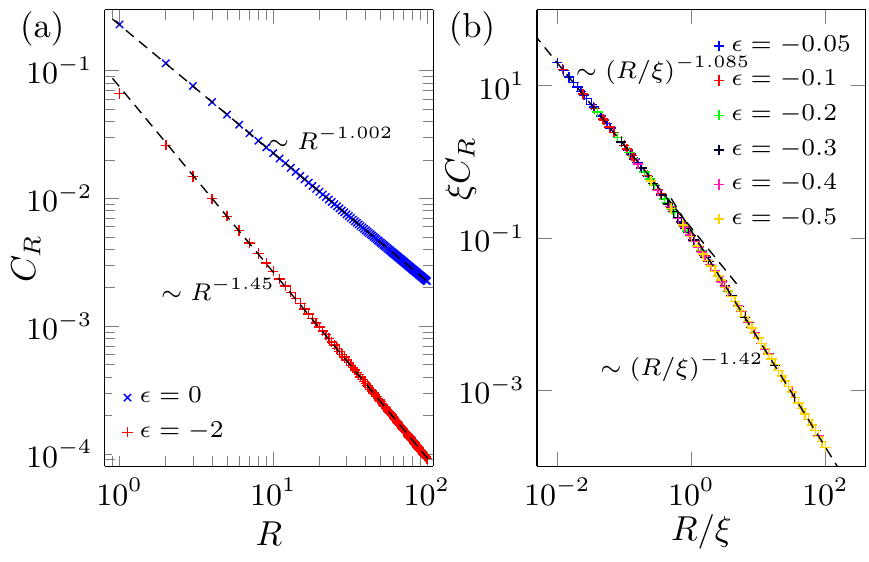}
\caption{  {\bf Long-range extended Ising model.}
In (a),
the static correlation function $C_R$ at the critical $\epsilon=0$ for $\alpha=3/2$. 
The solid line is a linear fit with a slope $-1.002\approx -1 $.
The same panel shows the same function far from the critical point at $\epsilon=-2$. 
The solid line is a linear fit with a slope $-1.45 \approx -(3 -\alpha) $.  
In (b),
the scaled static correlation function $\xi C_R$ as a function of scaled distance $R/\xi$.
In accordance with the static scaling hypothesis (\ref{staticSH}),
the plots for different $\epsilon$ collapse to a unique scaling function.
We can see a crossover from the critical $(R/\xi)^{-1}$ to the off-critical $(R/\xi)^{-3/2}$
near $R/\xi=1$. The straight lines are linear fits with slopes $-1.085\approx-1$ and $-1.42\approx -(3-\alpha)$ for smaller and larger $R/\xi$, respectively. 
}
\label{fig2}
\end{figure}

\begin{figure*}[t]
\includegraphics[width=\textwidth]{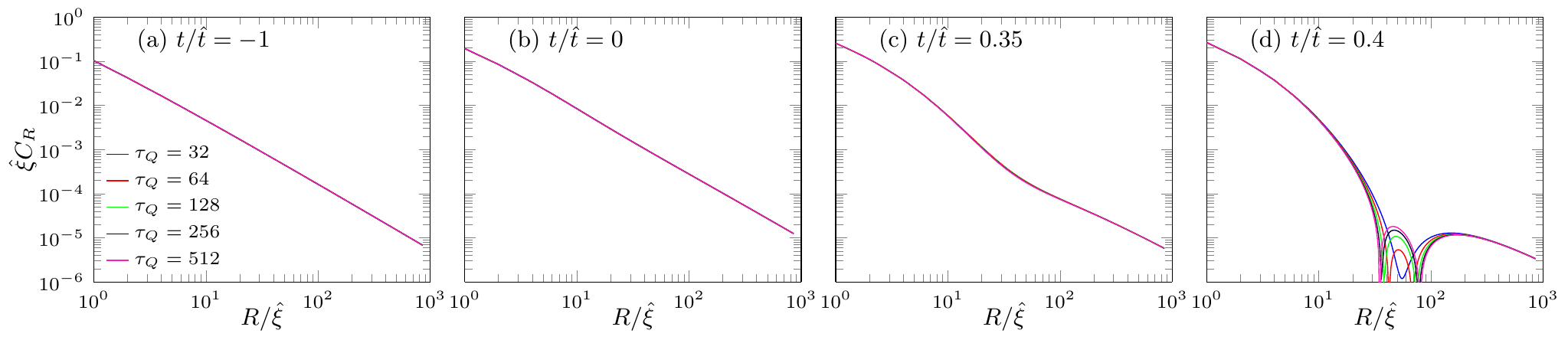}
\caption{ {\bf Long-range extended Ising model.}
The scaled dynamical correlation function $\hat\xi C_R$ for different quench times $\tau_Q$ at different scaled times, $s={t}/{\hat{t}}$, during the quench.
Here $\alpha=3/2$.
From left to right: $s=-1,0,0.35,0.4$. 
In (a) at $s=-1$,
the evolution is still adiabatic and the plots for different $\tau_Q$ collapse according to the static scaling hypothesis which is equivalent to the KZ hypothesis in this early adiabatic regime.
In (b) at $s=0$,
when crossing the critical point the evolution is diabatic. Different $\tau_Q$ collapse to a common scaling function $F(s,R/\hat\xi)$ in accordance with the KZ scaling hypothesis (\ref{CRscaling}).
In (c) at $s=0.35$,
the correlation function begins to bend down at distances between $R/\hat\xi=10$ and $100$.
Different $\tau_Q$ collapse, as the evolution is still in the KZ regime.
Finally in (d) at $s=0.4$,
the collapse fails due to the phase ordering after the system left the KZ regime and entered the adiabatic one. At the distances between $R/\hat\xi=10$ and $100$ the correlations become negative for longer $\tau_Q$.
}
\label{fig:corrp0m}
\end{figure*}

Thanks to the string operator in the long-range interaction terms, the model can be mapped to a quadratic free fermion model and solved analytically. After the Jordan-Wigner transformation (\ref{JordanWigner}) the Hamiltonian (\ref{Hamil_LR}) becomes:
\bea
H &=& -\sum_n \left(1-\epsilon\right)\left(c_nc_n^\dag-c_n^\dag c_n\right) \nb \\
&&- \sum_{n,r} J_r\left( c_n^\dag c_{n+r} + c_n^\dag c_{n+r}^\dag + {\rm h.c.}  \right),
\label{HcLR}
\eea
for the anti-periodic boundary conditions. This representation is also known as the long-range Kitaev model \cite{LRKitaev1, LRKitaev2}, where the hopping and pairing terms are of equal strength. After Fourier transformation (\ref{Fourier}),
\bea
H &=&
-2\sum_{k>0} 
(1-\epsilon -  \Re(\tilde{J_k})) \left(c_k^\dag c_k+c_{-k}^\dag c_{-k}\right) \nb \\
&&+  ~\Im(\tilde{J_k})\left(c_{k}^\dag c_{-k}^\dag + c_{-k}c_{k}\right),
\label{HkLR}
\eea
where  $\Re(\tilde{J_k})$ and  $\Im(\tilde{J_k})$ are, respectively, real and imaginary parts of the Fourier transform $\tilde{J_k}=\sum_r J_r e^{ikr}$. We have $\tilde{J_k}=\frac{{\rm Li}_{\alpha}(e^{ik})}{\zeta(\alpha)}$, where ${\rm Li}$ is the polylogarithm function: ${\rm Li}_\alpha(x)=\sum_{n=1}^\infty\frac{x^n}{n^\alpha}$.

We can now find the stationary Bogoliubov-de Gennes equations (\ref{XYBdGk}) with
$
A_k(\epsilon)=1-\epsilon-\Re(\tilde{J_k}),~~ B_k=\Im(\tilde{J_k})
$
and eigenfrequencies
$
\omega_k=2\sqrt{A_k^2(\epsilon)+B_k^2}.
$
We have a critical point at $\epsilon=0$ where the gap closes for $k=0$. Another critical point, 
not to be considered here, is at $\epsilon=2(1-2^{-\alpha})$ and $k=\pi$. The dispersion relation at the critical $\epsilon=0$ is
\bea
\omega_k \simeq |k|^{\alpha-1},
\eea
hence the dynamical exponent is $z=\alpha-1\in(0,1)$. On the other hand, for small $\epsilon$, the gap at $k=0$ closes as
$ 
\omega_0 = 2|\epsilon|,
$
hence $z\nu=1$.

In a short range model for large $R$ the correlation function decays exponentially with $R$ when the system is away from the critical point but this does not need to hold for long-range interactions. Indeed, our system has a power law scaling even far away from the critical point \cite{LRKitaev1,LRKitaev2}:
\bea 
C_R &=&
\langle \sigma_i^x \sigma_{i+1}^z \ldots \sigma_{i+R-1}^z \sigma_{i+R}^x\rangle
\sim 1/R^{3-\alpha},
\label{Corr}
\eea
compare Fig.~\ref{fig2}(a). On the other hand, as discussed in more detail in Ref.~\onlinecite{natun}, at the critical point we expect a critical power law $C_R \sim 1/r^{\eta}$ with $\eta<\alpha$. Indeed, in Fig.~\ref{fig2}(a) we find that $\eta=1$ for $\alpha=3/2$. Similarly as for short-range interactions, for a small $\epsilon$ we expect a cross-over between the two power laws when $R$ is close to $\xi\sim\epsilon^{-\nu}$. Note that here $\xi$ is not the usual exponential correlation length, even though it scales with $\epsilon$ in the characteristic way. The crossover can be verified with a static scaling hypothesis:
\bea 
\xi^\eta ~C_R = F_{\rm st}\left( R/\xi \right).
\label{staticSH}
\eea 
With $\xi=\epsilon^{-\nu}$ the plots of the scaled correlator $\xi^\eta C_R$ as a function of the scaled distance $R/\xi$ for different $\epsilon$ should collapse to a common static scaling function $F_{\rm st}(x)$. We expect $F_{\rm st}(x)$ to cross-over around $x=1$ between the critical tail $x^{-1}$ for small $x$ and $x^{3-\alpha}$ for large $x$. Indeed, this is what we see in Fig.~\ref{fig2}(b). The unscaled leading static tail is, therefore, 
\be 
C_R\simeq \xi^{1/2} R^{-3/2}
\label{lrtailstatic}
\ee 
for $R\gg\xi$.

\begin{figure}[b]
\vspace{-0.5cm}
\includegraphics[width=\columnwidth]{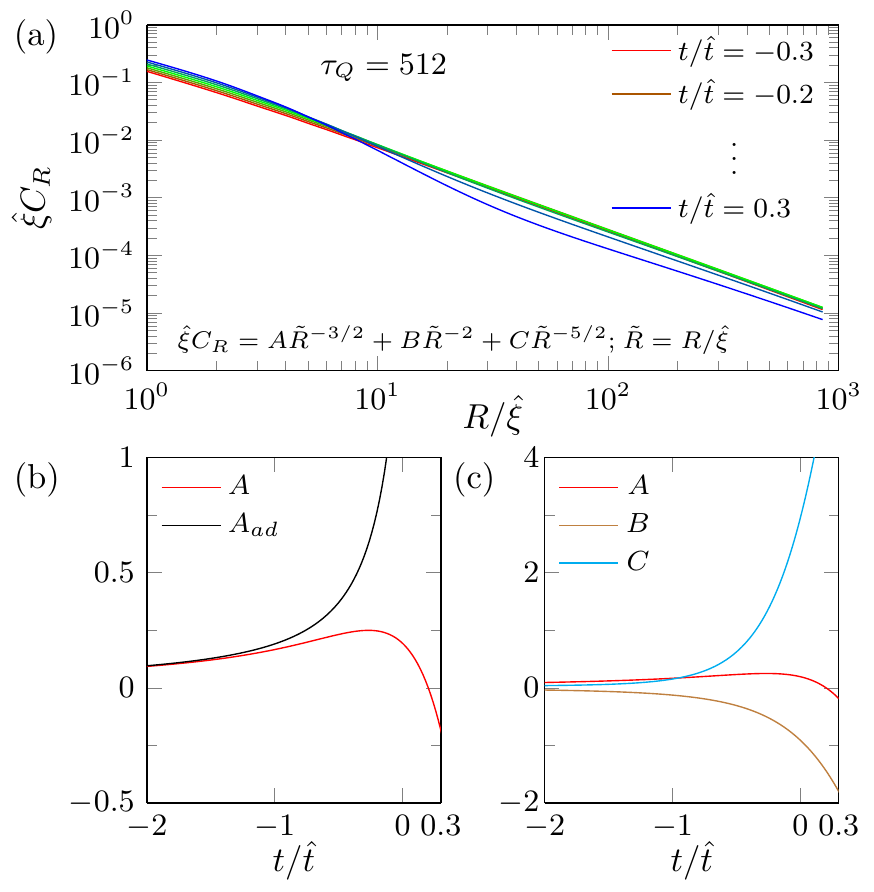}
\vspace{-0.5cm}
\caption{ {\bf Long-range extended Ising model.}
In (a), 
scaled correlator $\hat\xi C_R$ for $\alpha=3/2$ as a function of scaled distance for different scaled times in the log-log scale.
In (b),
in the inset the coefficients $A,B,C$ in Eq.~\eqref{ABC(s)fit} fitted to the tail of the scaled correlator are shown as a function of scaled time $s=t/\hat t$. The main picture shows $A(t/\hat t)$ together with the adiabatic $A_{\rm ad}(t/\hat t)$.
In the initial adiabatic stage $A$ follows $A_{\rm ad}$. 
In the KZ stage $A_{\rm ad}$ diverges to infinity at the critical point forcing $A$ to lag behind.
}
\label{fig:fig2a}
\end{figure}

Having verified the static hypothesis, we are encouraged to propose its dynamical version in the same form (\ref{CRscaling}) as for local interactions. The dynamics is described by the time-dependent Bogoliubov-de Gennes equations (\ref{tBdGk}) whose solution is presented in Appendix \ref{AppLR}. The KZ mechanism has recently been verified in a similar model~\cite{amitdutta}, where defect density was found to scale as $\tau_Q^{-1/2(\alpha-1)}$ in agreement with the KZ prediction $\tau_Q^{-\nu/(1+z\nu)}$, see also Ref.~\cite{KZLR2}. Further developments, like scaling of excitation energy (consistent with KZ) or possible experimental implementation, were considered in Refs.~\onlinecite{MorigiLR,LRPlenio}.
Here, we are interested in the build-up of correlations during the quench. In Fig.~\ref{fig:corrp0m}(a) we show scaled correlation functions at different scaled times. For definiteness, in this figure we set $\hat t=\tau_Q^{1/2}$ and $\hat\xi=\tau_Q^{1/2(\alpha-1)}$ with both numerical prefactors equal to $1$. We can see that up to $t=0.35\hat t$ the scaled correlators for different $\tau_Q$ collapse demonstrating the KZ scaling hypothesis in Eq.~\eqref{CRscaling}. At $t/\hat t\approx 0.4$ they already fail to collapse, signalling the end of the KZ regime and the beginning of phase ordering in the second adiabatic stage. 

In Fig.~\ref{fig:fig2a}(a), we collect scaled correlators for different scaled times, $s=t/\hat t$, in the KZ regime. In order to gain better insight, in Fig.~\ref{fig:fig2a}(b) we fit their tails with a function
\be 
\hat \xi C_R(s)=A(s)(R/\hat\xi)^{-3/2}+B(s)(R/\hat\xi)^{-2}+C(s)(R/\hat\xi)^{-5/2}.
\label{ABC(s)fit}
\ee 
Its form is motivated by correlation tails after a sudden quench considered in appendices \ref{SuddenAppInfinite} and \ref{SuddenAppFinite}. Here, the coefficients $A,B,C$ are functions of the scaled time shown in Fig.~\ref{fig:fig2a}(c). On top of the leading $A(s)$ tail, sub-leading dynamical correlations build up in time, with $B(s)$ growing negative and $C(s)$ positive. Their behavior is similar to what is observed after a sudden quench to the critical $\epsilon=0$ either from $\epsilon=-\infty$ or $\epsilon=-1$, see appendices \ref{SuddenAppInfinite} and \ref{SuddenAppFinite}, respectively. The main Fig.~\ref{fig:fig2a}(b) shows $A(s)$, i.e., the coefficient of the leading long-range tail $\propto R^{-3/2}$.

The same figure shows $A_{\rm ad}(s)$, i.e., the same coefficient but in case the evolution were adiabatic. $A_{\rm ad}(s)$ was obtained by a fit to the far tail of the static correlation function at $\epsilon(s)=s/(\tau_Q/\hat t)$. The same $A_{\rm ad}(s)$ can be obtained by equating the static tail (\ref{lrtailstatic}) with the adiabatic tail $C_R=A_{\rm ad}(s)\hat\xi^{1/2}/R^{3/2}$, compare (\ref{ABC(s)fit}):
\be 
A_{\rm ad}(s) \simeq
\left[\frac{\xi(s)}{\hat\xi}\right]^{1/2} \simeq
\left[\frac{\epsilon(s)^{-\nu}}{\hat\xi}\right]^{1/2} \sim
s^{-\nu/2}.
\ee 
$A_{\rm ad}(s)\sim s^{-1}$ for $\alpha=3/2$ is consistent with the divergence we observe in Fig.~\ref{fig:fig2a}(b). We can also see that, as expected, $A(s)$ follows $A_{\rm ad}(s)$ in the initial adiabatic stage up to $s\simeq -1$. After that, in the diabatic KZ stage, it lags behind as it cannot catch up with the diverging $A_{\rm ad}(s)$. If it did not, and the evolution were adiabatic, then the correlation range -- defined by $R$ where $C_R$ falls below a fixed small cutoff -- would diverge near the critical point as $|t|^{-2/3}$.

After $A(s)$ begins to lag behind the adiabatic evolution, it continues to grow at a finite rate until the critical point. This is the stage where the long-range model is the most reminescent of the standard causal KZ picture. After the critical point $A(s)$ begins to dip down around $s\approx 0.2$ but this is where the KZ scaling hypothesis (\ref{CRscaling}) begins to be violated, see Fig.~\ref{fig:corrp0m}(d).

\section{ Conclusion } 
\label{SectionConclusion}

We recall the causal/sonic horizon version of the Kibble-Zurek mechanism (KZM).
In the initial adiabatic stage of the evolution, 
the correlation range follows its adiabatic counterpart.
At $-\hat t$ it begins to lag behind the diverging adiabatic range and continues to grow at a finite rate set by the speed limit $\simeq \hat\xi/\hat t$.
This way, between $-\hat t$ and $+\hat t$, the correlation range can increase several times at odds with the impulse approximation where it remains frozen. 

There are notable exceptions, like the random quantum Ising chain, where the increase is small due to localization of excited quasiparticles that prevents them from spreading entanglement. There are interesting generalizations, like the long-range extended Ising model, where it is the power law correlation tail, rather than the exponential correlation length, that lags behind its diverging adiabatic evolution.

\acknowledgements
AS would like to thank Titas Chanda for useful discussions.
We acknowledge funding by National Science Centre (NCN), 
Poland under projects No.~2016/23/B/ST3/00830 (JS,AF) 
and No. 2016/23/D/ST3/00384 (MMR), 
NCN together with European Union through QuantERA ERA NET 
program No.~2017/25/Z/ST2/03028 (AS,DS,JD),
and Department of Energy under the Los Alamos National Laboratory 
LDRD Program (WHZ).
WHZ was also supported by the U.S. Department of Energy, Office of Science, Basic Energy
Sciences, Materials Sciences and Engineering Division, Condensed Matter Theory Program.
AF acknowledges financial support by Polish Ministry of Science and Education, project No. DI2015 021345, from the budget funds for science in 2016-2020 under the "Diamond Grant" program.
This research was carried out with the equipment purchased thanks to the financial support of the European Regional Development Fund in the framework of the Polish Innovation Economy Operational Program (contract No. POIG.02.01.00-12-023/08). 


\appendix

\section{Jordan Wigner, Fourier and Bogoliubov transformations}

After the Jordan-Wigner transformation,
\bea
\sigma^x_n &=&- \left( c_n + c_n^\dagger \right)
 \prod_{m<n}(1-2 c^\dagger_m c_m)~, \nonumber \\
\sigma^y_n &=& i \left( c_n - c_n^\dagger \right)
 \prod_{m<n}(1-2 c^\dagger_m c_m)~,\label{JordanWigner} \\
\sigma^z_n &=& 1~-~2 c^\dagger_n  c_n~ \nonumber, 
\eea
introducing fermionic operators $c_n$ that satisfy
$\left\{c_m,c_n^\dagger\right\}=\delta_{mn}$ and 
$\left\{ c_m, c_n \right\}=\left\{c_m^\dagger,c_n^\dagger \right\}=0$
the Hamiltonian $H$ becomes~\cite{Schultz1964}
\be
 H~=~P^+~H^+~P^+~+~P^-~H^-~P^-~.
\label{Hc}
\ee
Above
$
P^{\pm}=\frac12\left[1\pm P\right]
$
are projectors on subspaces with even ($+$) and odd ($-$) parity,
\be 
P~=~
\prod_{n=1}^N\sigma^z_n ~=~
\prod_{n=1}^N\left(1-2c_n^\dagger c_n\right), ~
\label{P}
\ee
and  
$
H^{\pm}
$
are corresponding reduced Hamiltonians. The $c_n$'s in $H^-$ satisfy
periodic boundary condition $c_{N+1}=c_1$, but the $c_n$'s in $H^+$
are anti-periodic, $c_{N+1}=-c_1$.

The initial ground state at $\epsilon\to-\infty$ has even parity, 
hence we can confine to the even subspace. 
The translationally invariant $H^+$ is diagonalised by a Fourier transform 
followed by a Bogoliubov transformation \cite{Schultz1964}. 
The anti-periodic Fourier transform is  
\be
c_n~=~ 
\frac{e^{-i\pi/4}}{\sqrt{N}}
\sum_k c_k e^{ikn}~,
\label{Fourier}
\ee
where the pseudomomentum takes half-integer values
\be
k~=~
\pm \frac12 \frac{2\pi}{N},\pm\frac32 \frac{2\pi}{N},
\dots,
\pm \frac{N-1}{2} \frac{2\pi}{N}~.
\label{halfinteger}
\ee
Diagonalization of $H^+$ is completed by a Bogoliubov transformation,
\be
c_k ~=~ U_k  \gamma_k + V_{-k}^*  \gamma^\dagger_{-k},
\label{Bog}
\ee
provided that Bogoliubov modes $(U_k,V_k)$ are eigenstates of stationary
Bogoliubov-de Gennes equations with positive eigenfrequency $\omega_k$.

\section{Landau-Zener model}
\label{LZ}

The canonical LZ model is
\be
i\frac{d}{dt'}
\left(
\begin{array}{c}
u_k \\
v_k
\end{array}
\right)=
\frac12
\left[
-\frac{t'}{\tau_k}\sigma^z+\sigma^x
\right]
\left(
\begin{array}{c}
u_k \\
v_k
\end{array}
\right),
\label{tBdGkLZ}
\ee
where $\tau_k$ is a transition time.
Its solution is
\bea
u_k &=&
e^{-\frac{\pi}{16}\tau_k}
D_{\frac14i\tau_k}(z)
e^{i\pi/4},\\
v_k &=&
\frac12
e^{-\frac{\pi}{16}\tau_k}
D_{-1+\frac14i\tau_k}(z)
\sqrt{\tau_k}.
\label{general} 
\eea
Here $D_m(z)$ is the Weber function with an argument
\be 
z=e^{3\pi i/4}\frac{t'}{\sqrt{\tau_k}},
\ee 
see for instance Ref.~\onlinecite{Damski_PRA_2006}. 

\section{Time-dependent extended XY model }
\label{XYtime}

The time-dependent problem is
\bea
i\frac{d}{dt}
\left(
\begin{array}{c}
u_k \\
v_k
\end{array}
\right)=
2\left[\sigma^z A_k(t/\tau_Q)+\sigma^x B_k \right] 
\left(
\begin{array}{c}
u_k \\
v_k
\end{array}
\right).
\label{tBdGk}
\eea
It can be mapped to the canonical LZ model where 
\bea 
\tau_k &=& 4\tau_Q(a\gamma\sin k+b\delta\sin 2k)^2\, \\
    t' &=& 4\tau_Q(a\gamma\sin k+b\delta\sin 2k)\times \nb\\
       & & \left(\frac{t}{\tau_Q}-(1-a\cos k - b\cos 2k)\right).
\eea
Only small quasimomenta up to $\hat k\simeq\tau_Q^{-1/6}$ become excited, hence for $k\ll1$ we can approximate: 
\bea
u_k &=&
e^{-\frac14\pi q^2}
D_{iq^2}(z)
~e^{i\pi/4},\nonumber\\
v_k &=&
e^{-\frac14\pi q^2}
D_{-1+iq^2}(z)
~2q,\nonumber\\
z &=&
4e^{3\pi i/4}\left(2s-\frac{q^{4/3}3^{1/3}}{\tau_Q^{1/6}}\right),
\label{rescuv} 
\eea
where
$
q = \frac13 k^3 \sqrt{\tau_Q}
$
is a scaled quasimomentum and 
$ 
s=t/\sqrt{\tau_Q}
$ 
is a scaled time. These formulas are consistent with
$\hat\xi\simeq\tau_Q^{1/6}$ and $\hat t\simeq\tau_Q^{1/2}$, respectively.

Only $q$ up to $q\approx1$ get excited. For them, when $\tau_Q$ is large enough,
we can further approximate 
\be 
z=4e^{3\pi i/4}\left(2s-\frac{q^{4/3}3^{1/3}}{\tau_Q^{1/6}}\right)\approx 8e^{3\pi i/4}s
\label{approxzXY}
\ee 
and obtain
\bea
u_k &=&
e^{-\frac14\pi q^2}
D_{iq^2}\left(8e^{3\pi i/4}s\right)
~e^{i\pi/4},\nonumber\\
v_k &=&
e^{-\frac14\pi q^2}
D_{-1+iq^2}\left(8e^{3\pi i/4}s\right)
~2q.
\label{infuv} 
\eea
In accordance with the KZ scaling hypothesis (\ref{CRscaling}),
these non-adiabatic modes depend on the scaled variables $s$ and $q$ only. 
However, the approximation in (\ref{approxzXY}) requires $\tau_Q^{1/6}\gg 1$ -- which is why we consider large $\tau_Q\simeq 2^{20}$ in Fig.~\ref{fig:XYspeed}.


\section{Solving the time-dependent Bogoliubov-de Gennes equations}
\label{tB-deG_RIM}

The time-dependent problem in Eq.~(\ref{t_B-deG}) is 

\bea
\centering
i\frac{du_{nm}^{\pm}}{dt} = 2(h_{n}-\epsilon(t))u_{nm}^{\mp}-2J_{n}u_{n\mp1,m}^{\mp}
\label{t_B-deG_App}
\eea

The central theme in the time-evolution of quantum mechanical states is that of exponentiation of the Hamiltonian which can be attributed to the unitary nature of its dynamics. This becomes notoriously hard when the Hamiltonian consists of non-commuting terms. To add to that, the Random Ising Model also lacks translational invariance. It renders useless the mapping of  time-dependent  Bogoliubov-de Gennes equations to the canonical LZ model---unlike the other models used in this article. Consequently we resort to techniques like the Suzuki-Trotter (ST) expansion of the exponential which has the desirable feature of conserving the norm of the evolution. The ST approximation deals with approximating the exponential, expressed in terms of the two non-commuting pieces, by a product of exponential. ST of second order can be written as 
\bea
e^{\delta(A+B)} = \lim_{\delta \to 0}  e^{\frac{\delta}{2}A} e^{\delta B}e^{\frac{\delta}{2}A} + \mathcal{O}(\delta)^3 \nonumber
\label{sz2}
\eea
We separate the field part (A) and the coupling part (B) and solve them separately for a small time $dt$.
\begin{eqnarray*}
\text{field:\hspace*{0.05cm}}\left( \begin{matrix} u_{n}^{+}(t+dt) \\ u_{n}^{-}(t+dt) \end{matrix} \right) &=& e^{-i A(t+dt/2) dt}\left( \begin{matrix} u_{n}^{+}(t) \\ u_{n}^{-}(t) \end{matrix} \right), \\
\text{coupling:\hspace*{0.05cm}}\left( \begin{matrix} u_{n+1}^{+}(t+dt) \\ u_{n}^{-}(t+dt) \end{matrix} \right) &=& e^{-i B dt}\left( \begin{matrix} u_{n+1}^{+}(t) \\ u_{n}^{-}(t) \end{matrix} \right),
\end{eqnarray*}
where $A(t) = (h_{n}-\epsilon(t))\sigma_{x} $ and $B =-2J_{n}\sigma_{x}$, and drop the index $m$ for convenience.
We begin our quench at $\epsilon(t) = 5$.


\section{Calculation of Two-Site Correlation function in the Random Ising Model}
\label{appcorrRIM}
Let us define two operators $a_{i}$ and $b_{i}$ in terms of fermionic operators as $a_{i}=c_{i}^{\dag}+c_{i}$ and $b_{i}=c_{i}^{\dag}-c_{i}$.
Using Jordan-Wigner transformation(see \ref{JordanWigner}) and equation(\ref{bogoliubov_RIM}), the two site correlation function along the longitudinal direction can be written as  
\bea
C_{R}^{xx} &=& \langle\Psi(t)|\sigma_{i}^{x}\sigma_{i+R}^{x}|\Psi(t)\rangle\\ &=& \langle b_{i}a_{i+1}b_{i+1}a_{i+2}...a_{i+R-1}b_{i+R-1}a_{i+R}\rangle
\label{corrxxab}
\eea
where $|\Psi(t)\rangle$ is the evolving state. 
Our final longitudinal correlation looks like 
\bea
\label{corrRIMpffafian}
\mid C^{xx}_{R} \mid =  \sqrt{\mid {\rm det}(A_{R}^{xx})\mid},
\eea
where 
\bea
\label{enlargedcorrRIM}
A_{R}^{xx} = \begin{bmatrix} \langle a_{i+1}a_{j+1}\rangle & \langle b_{i}a_{j+1} \rangle \\
\langle a_{i+1}b_{j} \rangle & \langle b_{i}b_{j} \rangle
\end{bmatrix}_{i,j=1,...,R}
\eea
and where we redefine $\langle a_{i}a_{i} \rangle = \langle b_{i}b_{i} \rangle = 0$. 
Eqs.~\eqref{corrRIMpffafian} and \eqref{enlargedcorrRIM} follows from \eqref{corrxxab} 
using the Wick theorem. See for instance Ref.~\onlinecite{youngRIM-a, youngRIM-b} for more details on similar calculations. Finally, two-point correlation functions are obtained as 
\bea 
\langle a_{i}a_{j}\rangle&=&\sum_{m}u_{im}^{+}u_{jm}^{+*}, \nonumber\\ 
\langle a_{i}b_{j}\rangle&=&\sum_{m}u_{im}^{+}u_{jm}^{-*}, \nonumber\\
\langle b_{i}a_{j}\rangle&=&-\sum_{m}u_{im}^{-}u_{jm}^{+*}, \nonumber\\
\langle b_{i}b_{j}\rangle&=&-\sum_{m}u_{im}^{-}u_{jm}^{-*}. \nonumber
\label{abinuv}
\eea
where $\sum_{m}$ denotes the summation over all Bogoliubov modes.

\section{Time-dependent long-range Ising model }
\label{AppLR}

The time-dependent Bogoliubov-de Gennes equations map to the canonical LZ model (\ref{tBdGkLZ})
where 
\bea 
&& \tau_k = 4\tau_Q\Im(\tilde{J_k})^2\,\nb \\
&& t' = 4\tau_Q~\Im(\tilde{J_k})\left(\frac{t}{\tau_Q}-1+\Re(\tilde{J_k})\right)\nb \\
\eea
For the relevant $k\ll1$, its exact solution can be approximated by
\bea
u_k &=&
e^{-\frac14\pi q^2}
D_{iq^2}(z)
~e^{i\pi/4},\nonumber\\
v_k &=&
e^{-\frac14\pi q^2}
D_{-1+iq^2}(z)
~q,\nonumber\\
z &=&
2e^{3\pi i/4}\left(s-C\frac{q}{\tau_Q}\right),~ 
\label{rescuvLR} 
\eea
where $C\simeq 1$ is a constant, 
$
q \simeq {k}^{1/2} {\tau_Q}^{1/2} 
$
is a scaled quasi-momentum and 
$ 
s=t/\sqrt{\tau_Q}
$ 
a scaled time. These formulas are consistent with the KZ scales $\hat t\simeq\tau_Q^{1/2}$ and
$\hat\xi \simeq \tau_Q$ for $\alpha=3/2$. When $\tau_Q\gg1$ we can approximate 
$z\approx 2e^{3\pi i/4}s$ and the solution depends on the scaled variables only.
In this regime we also expect
\be 
\hat v=\hat\xi/\hat t\simeq \tau_Q^{1/2}.
\ee
as a relevant velocity of quasiparticles excited during the quench.

With the exact solution, one can calculate the correlation function,
\bea
C_R = \delta_{R,0}-2\alpha_R+2\Re(\beta_R), 
\label{corralbe}
\eea
where $\alpha_R=\frac{1}{\pi}\int_0^\pi dk~|u_k|^2\cos{kR}$ and 
$\beta_R=\frac{1}{\pi}\int_0^\pi dk~ u_k v_k^*\sin{kR}$.

\section{
Infinite sudden quench from $\epsilon=-\infty$ to the critical point in the long-range Ising model
}
\label{SuddenAppInfinite}

In this and the next appendix, we consider sudden quenches that have also been considered before in the context of dynamical phase transitions in the Lochschmidt echo\cite{LREnss}. At $\epsilon=-\infty$ (corresponding to infinite transverse field) the ground state---with all spins pointing in $+x$ direction---has no correlations. The correlations begin to build up after a sudden quench at $t=0$ from $\epsilon=-\infty$ to the critical $\epsilon=0$. The initial Bogoliubov modes at $t=0$ are 
\be
\left[
\begin{array}{c}
u_k(0) \\
v_k(0)
\end{array}
\right] = 
\left[
\begin{array}{c}
1\\
0
\end{array}
\right].
\ee 
After the quench they evolve as
\bea
\left[
\begin{array}{c}
u_k(t) \\
v_k(t)
\end{array}
\right]&=&
\left[
\begin{array}{c}
U_k^2 e^{-i\omega_k t} + V_k^2 e^{i\omega_k t}\\
U_k V_k (e^{-i\omega_k t}-e^{i\omega_k t}) 
\end{array}
\right].
\label{suddenBogo}
\eea
Here $\omega_k$ are the frequencies and $(U_k,V_k)$ the stationary Bogoliubov modes at $\epsilon=0$. With this solution we can now calculate the correlation function in (\ref{corralbe}) consisting of two terms: $\alpha_R$ and $\beta_R$.

\begin{figure}[t]
\includegraphics[width=0.98\linewidth]{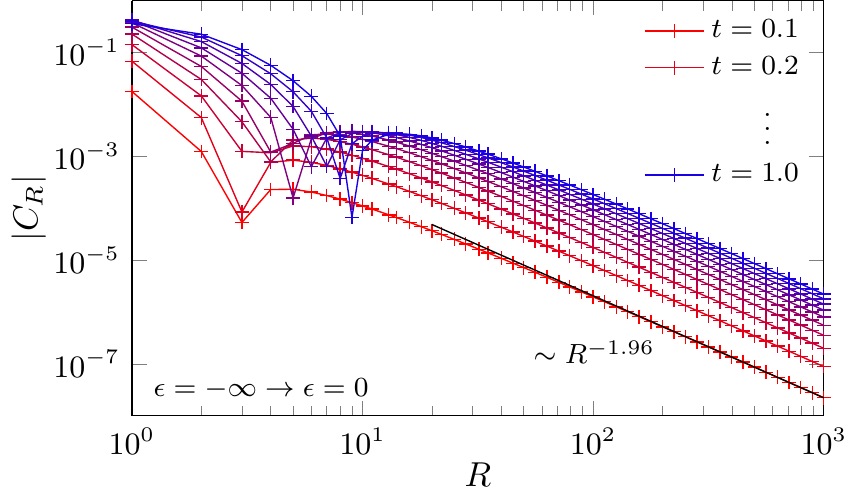}
\caption{
{\bf Long-range extended Ising model---sudden quench.}
Spreading of correlation after a sudden quench from $\epsilon=-\infty$ to $\epsilon=0$.  
The tail of $C_R$ can be fitted with $R^{-1.96}$ in good agreement with the predicted $R^{-2}$.
$C_R$ changes sign from positive for small $R$ to negative for large $R$
as manifested by the kinks in the plot of $|C_R|$ versus $R$. 
}
\label{fig:appen1}
\end{figure}

Let us focus first on $\alpha_R$. It is useful to notice that 
\bea
|u_k(t)|^2&=&1+2U_k^2V_k^2(\cos{2\omega_k t}-1),\nonumber \\
&=& 1 +  2U_k^2V_k^2 \sum_{n=1}^\infty (-1)^n \frac{(2\omega_k t)^{2n}}{(2n)!} 
\label{abar}
\eea
Correlations at the largest distances $R$ are provided by pairs of correlated quasiparticles with the largest velocities. From the spectrum $\omega_k$ for $1<\alpha<2$ it can be readily seen that the fastest quasiparticles are those with $k \to 0$. Their velocity tends to infinity. For small $k$ we can approximate \cite{NISThandbook}
\bea
\Im(\tilde{J_k}) &\approx& A_\alpha k^{\alpha-1} + C_\alpha k, \nonumber\\
\Re(\tilde{J_k}) &\approx& B_\alpha k^{\alpha-1} + 1, 
\eea
where $A_\alpha = \frac{\Gamma(1-\alpha)\cos{(\pi\alpha/2)}}{\zeta{(\alpha)}}$, $B_\alpha = \frac{\Gamma(1-\alpha)\sin{(\pi\alpha/2)}}{\zeta{(\alpha)}}$ and $C_\alpha = \frac{\zeta(\alpha-1)}{\zeta(\alpha)}$. 
Using these approximations we obtain the following asymptotes: 
\bea
\omega_k   &\approx & 2 \frac{\Gamma(1-\alpha)}{\zeta(\alpha)}k^{\alpha-1},\nonumber \\
U_k^2V_k^2 &\approx & C_1^\alpha (1+C_2^\alpha k^{2-\alpha}), 
\label{uvsq}
\eea
with $C_1^\alpha = \frac14 \frac{A_\alpha^2}{A_\alpha^2+B_\alpha^2}$ and $C_2^\alpha = 2(\frac{C_\alpha}{A_\alpha} - \frac{A_\alpha C_\alpha}{A_\alpha^2+B_\alpha^2})$. 
Inserting (\ref{abar}) and (\ref{uvsq}) we obtain
\bea
\alpha_R&\approx &\frac{1}{\pi}\int_0^\pi dk~|u_k(t)|^2\cos{kR} \nonumber\\
&\approx& \delta_{0,R} + \frac{2}{\pi} \sum_{n=1}^\infty (-1)^n \frac{(4\frac{\Gamma(1-\alpha)}{\zeta(\alpha)}t)^{2n}}{(2n)!}\times \nonumber \\
&\Big[& C_1^\alpha \int_0^\pi k^{2n(\alpha -1)} \cos{(kR)} dk \nonumber \\
&+& C_1^\alpha C_2^\alpha \int_0^\pi k^{2-\alpha+2n(\alpha -1)} \cos{(kR)} dk   \Big]
\label{aldekh}
\eea

\begin{figure}[t]
\includegraphics[width=\linewidth]{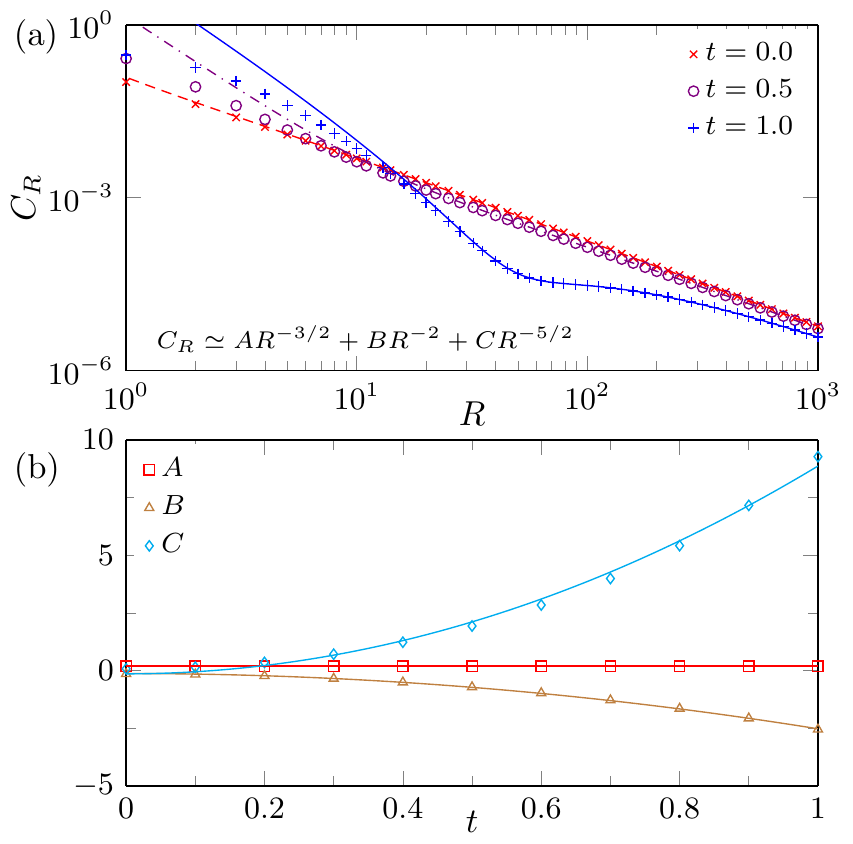}
\caption{{\bf Long-range extended Ising model---sudden quench.}
In (a),
evolution of the correlations (dots) after a finite sudden quench from $\epsilon = -1$ to the critical $\epsilon=0$. The correlators are fitted with the formula (\ref{fitABC}) in the range $900<R<1000$. The fits (solid lines) are plotted in the full range of $R$.  
In (b),
the coefficients $A,B,C$ in Eq.~\eqref{fitABC} as a function of time (dots).
Their time-dependence is fitted with parabolas $c_0+c_2 t^2$ (solid lines).
}
\label{fig:appen2}
\end{figure}

This approximation is valid for large $R$ where the integrals are dominated by small $k$. For small $k$ the integrand of the second integral is negligible as compared to the first one. By the same token we can neglect all terms except for $n=1$. For $n=1$ the first integral is
\bea
&&\int_0^\pi k^{2(\alpha -1)} \cos(kR)dk =\nonumber \\
&&
\frac{\pi^{2\alpha-1}}{2\alpha-1} 
{_1}{\rm F}{_2}\Big[\alpha-\frac12;\frac12,\alpha+\frac12;-\frac{\pi^2R^2}{4}\Big] 
\approx \nonumber\\
&&-\frac{\Gamma(2\alpha-1)\sin(\pi(\alpha-1))}{R^{2\alpha-1}}. \label{etaki}
\eea
Here ${_1}{\rm F}{_2}$ is the generalized Hypergeometric function~\cite{Abramowitz} and the last approximation is valid for large integer $R$. Combining (\ref{aldekh}) and (\ref{etaki}) we obtain 
\bea
\alpha_R &\approx& 
- \frac{2}{\pi}
  C_1^\alpha
  \frac{{\Gamma[1-\alpha]}^2}{{\zeta(\alpha)}^2}
  (1+\cos\pi\alpha) 
  \frac{t^2}{R^{2\alpha-1}}.
\eea
For $\alpha=3/2$ we have $\alpha_R\sim\frac{t^2}{R^2}$. 

Performing similar analysis, for $\alpha=3/2$ we obtain $\Re(\beta_R)\sim\frac{1}{R^{5/2}}$ which decays faster than $\alpha_R$. Therefore, in the correlation function (\ref{corralbe}), it is $\alpha_R$ that dominates the tail: $C_R\sim\frac{t^2}{R^{2}}$. In Fig.~\ref{fig:appen1} we show the numerically exact correlation function to confirm the validity of this asymptotic result for large $R$.

Before the infinite sudden quench, there is no static correlation tail, $\sim R^{-3/2}$, to be frozen into the state after the quench. All correlations have to build up dynamically by spreading of entangled pairs of quasiparticles with opposite quasimomenta excited by the sudden quench. The dynamical correlations develop a negative leading power-law tail $\propto R^{-2}$ and a positive next-to-leading term $\propto R^{-5/2}$. In the next appendix, we will see that after a finite quench, similar dynamical correlations build-up on top of a frozen static pre-quench tail $\propto R^{-3/2}$.

\section{Finite sudden quench from $\epsilon = -1$ to the critical point in the long-range Ising model}
\label{SuddenAppFinite}

In the previous appendix, the initial state before the quench was a product state with no correlations. Here, we consider a sudden quench from an initial state at a finite $\epsilon=-1$ where, for $\alpha=3/2$, the static correlation function decays like $R^{-3/2}$ for large $R$. Figure~\ref{fig:appen2}(a) shows the correlation function at different times after the sudden quench. In Fig.~\ref{fig:appen2}(b) we fit its tail with
\be 
C_R(t)=A(t)R^{-3/2}+B(t)R^{-2}+C(t)R^{-5/2}
\label{fitABC}
\ee
that includes also two sub-leading terms. $A(t)$ changes negligibly with respect to its pre-quench value, comparing to $B(t)$ and $C(t)$. The leading pre-quench static tail $\propto R^{-3/2}$ remains frozen after the sudden quench. 

The build-up of dynamical correlations is captured by the sub-leading terms. Their initial values are small corrections to the leading pre-quench static correlator. With time, they evolve away from the initial values like $\propto t^2$. Similarly as for the infinite sudden quench, $B(t)$ grows negative and $C(t)$ positive. Consequently, the dynamical part on top of the frozen static tail grows positive for small and negative for large $R$, see Fig.~\ref{fig:appen2}(a). 



\bibliographystyle{apsrev4-1}
\bibliography{KZhatv.bib}



\end{document}